\documentclass[%
 reprint,
 amsmath,amssymb,
 aps,
pre,
]{revtex4-1}

\usepackage{graphicx}% Include figure files
\usepackage{dcolumn}% Align table columns on decimal point
\usepackage{bm}% bold math
\usepackage{subeqnar}
\usepackage{overpic}
\usepackage{color}
\usepackage{adjustbox}
\usepackage[caption=false]{subfig}
\usepackage{rotating}
\usepackage{cleveref}

\crefformat{section}{\S#2#1#3} % see manual of cleveref, section 8.2.1

\begin{document}

%\title{Relating Convective, Mixing, and Chemical Time Scales to Rotating Detonation Wave Dynamics and Thermodynamic Power Cycles}
\title{Modeling Thermodynamic Trends of Rotating Detonation Engines}

\author{James Koch}
\email{james.koch@austin.utexas.edu}
\affiliation{Oden Institute for Computational Engineering and Sciences, University of Texas, Austin, TX, USA}
%\author{Mitsuru Kurosaka}
%\affiliation{William E. Boeing Department of Aeronautics and Astronautics, University of Washington, Seattle, USA}
%\author{Carl Knowlen}
%\affiliation{William E. Boeing Department of Aeronautics and Astronautics, University of Washington, Seattle, USA}
\author{J. Nathan Kutz}
\affiliation{Department of Applied Mathematics, University of Washington, Seattle, WA, USA}

\begin{abstract}
The formation of a number of co- and counter-rotating coherent combustion wave fronts is the hallmark feature of the Rotating Detonation Engine (RDE). The engineering implications of wave topology are not well understood nor quantified, especially with respect to parametric changes in combustor geometry, propellant chemistry, and injection and mixing schemes. In this article, a modeling framework that relates the time and spacial scales of the RDE to engineering performance metrics is developed and presented. The model is built under assumptions of backpressure-insensitivity and nominally choked gaseous propellant injection. The Euler equations of inviscid, compressible fluid flow in one dimension are adapted to model the combustion wave dynamics along the circumference of an annular-type rotating detonation engine. These adaptations provide the necessary mass and energy input and output channels to shape the traveling wave fronts and decaying tails. The associated unit processes of injection, mixing, combustion, and exhaust are all assigned representative time scales necessary for successful wave propagation. We find that the separation, or lack of, these time scales are responsible for the behavior of the system, including wave co- and counter-propagation and bifurcations between these regimes and wave counts. Furthermore, as there is no imposition of wave topology, the model output is used to estimate the net available mechanical work output and thermodynamic efficiency from the closed trajectories through pressure-volume and temperature-entropy spaces. These metrics are investigated with respect to variation in the characteristic scales for the RDE unit physical processes.
\end{abstract}
\maketitle

\section{Introduction}

The {\em Rotating Detonation Engine} (RDE) is a type of combustor with applications in aerospace propulsion and land-based power generation. The RDE uses detonative heat release (a nearly constant-volume process) as the dominant mode of energy addition to the fluid flow, contrasting deflagration-based, constant-pressure heat addition typical of aerospace engines. The engineering advantages of the RDE include a potential for greater thermal efficiency \cite{Nordeen2014,Shao2010}, the reduction of pumping requirements for propellant \cite{Sousa2017,Rankin2017}, mechanical simplification, and extraordinarily wide operability limits \cite{Anand2016,Fotia2016}. The quasi-steady state of the system is the saturation of a highly nonlinear combustion instability associated with the periodic geometry \citep{Anand2019}, leading to a number of coherent combustion wave fronts consuming propellant as they propagate around the combustor. These nonlinear dynamics of the combustion waves are fundamental to the operation of the RDE, unlike those of of conventional aerospace engines where such dynamics are suppressed. However, these dynamics are not well understood, especially with respect to relevant performance metrics and physics specific to particular engines, such as their geometry and their injection and mixing schemes. In this paper, we establish a link between the nonlinear dynamics of the rotating detonation waves, mechanical work output, and thermodynamic efficiency. This link is made possible through a modeling framework that blends a lumped-volume combustor model with the one-dimensional Euler equations of motion for an inviscid reacting flow. This framework provides input and output energy and mass flow channels required to shape the combustion wave fronts into a topology that satisfies the global input/output energy balance. This final topology includes co- and counter-rotating waves of varying number, plane waves, and stationary planar fronts.

\subsection{Experiments, Geometry, and Dynamics} \label{sec:background}
\begin{figure*}[]
        \centering
        \begin{overpic}[width=1.0\linewidth]{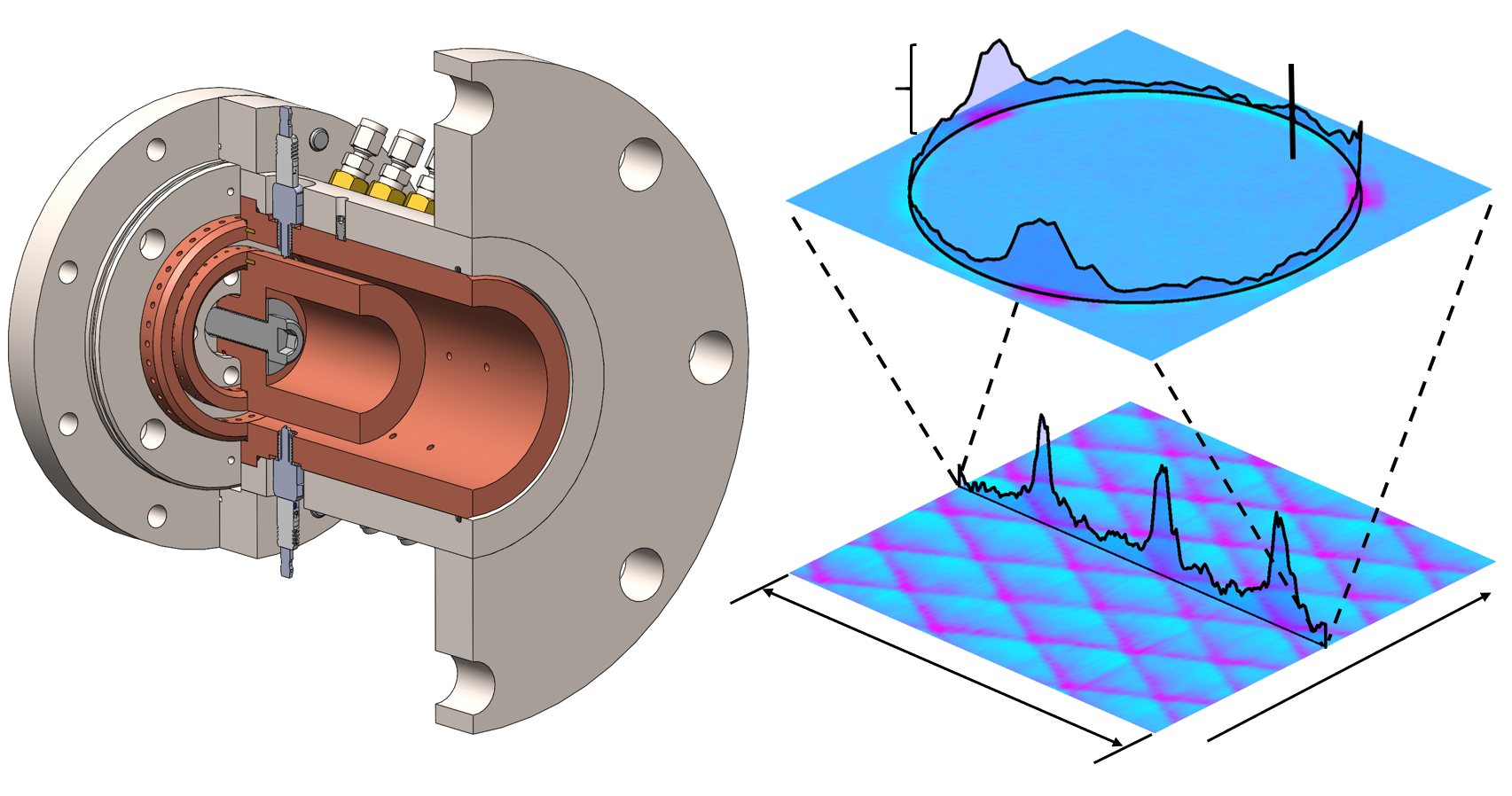}   
        \thicklines
                 
     	\put(55,5){$\theta \in [0,2\pi)$}                   
     	\put(88,5){Increasing time}                   
     	\put(47,48.5){Integrated}                   
        \put(47,46.5){pixel intensity}
        \put(47,44.5){(visible)}

        \put(87,47){$\theta = 0, 2\pi$}
        \put(71,40.5){\color{white}{High-speed}}
        \put(70,38.5){\color{white}{camera frame}}

		\put(23,16.5){\vector(0,1){6.5}}
        \put(20,15){Annular}
        \put(20,13){combustion}
        \put(20,11){chamber}

		\put(5,40){\vector(2,-1){6.5}}
        \put(0,44){Oxidizer}
        \put(0,42){injection}
        \put(0,40){ring}        
        
		\put(7,16.5){\vector(1,2){5.5}}
        \put(5,15){Fuel}
        \put(5,13){injection}
        \put(5,11){ring}
        
 		\put(22,46){\vector(-1,-1){2.5}}
        \put(20,49){Igniter}
        \put(18,47){(spark plug)}
        
        \put(34.5,20){\vector(-2,1){5.5}}
        \put(34.5,20){\vector(-1,4){2.0}}
        \put(35,19){Instrumentation}
        \put(39,17){ports}
		
        \put(2,50){(a)}
        \put(52,41){(b)}
        \put(52,17){(c)}

	    \end{overpic}  
	    \caption{A section view of the rotating detonation engine used for this study is shown in (a). Fuel and oxidizer are injected radially into the annular combustion chamber via concentric injector rings. Oxidizer is injected through the outside ring while fuel is injected through the interior ring. All experiments presented in this paper were performed with this injector hardware configuration. Each ring has 36 orifices. The fuel and oxidizer orifices are offset such that they are staggered; this is to promote vortical mixing rather than impingement-dominated mixing. Ignition is provided with an array of automotive spark plugs. Static pressure along the axis of the engine is measured with an array of pressure transducers to record the mean (time-averaged) operating conditions. The engine is mounted to a backpressure-controlled dump volume. Downstream exhaust routing allows for direct optical access of the annulus. For each experiment, the complete space-time history of the detonation waves is recorded with a high-speed camera. A high-speed camera frame from an experiment is shown in (b) with the integrated pixel intensity of the combustion chamber plotted on top of the annulus location. Each video frame can be recast as a column vector of pixel intensities that can be stacked to form a $\theta-t$ diagram of wave kinematics (c). In this experiment, six waves co-exist in the engine. Three travel clockwise and three travel counter-clockwise. Their kinematic traces leave a distinct and repeatable spatiotemporal pattern. The integrated luminosity trace in (b) and in (c) corresponds to the moment of collision between the co- and counter-rotating waves.}
		\label{fig:intro}
\end{figure*}

\begin{figure*}[t]
        \centering
        \begin{overpic}[width=1\linewidth]{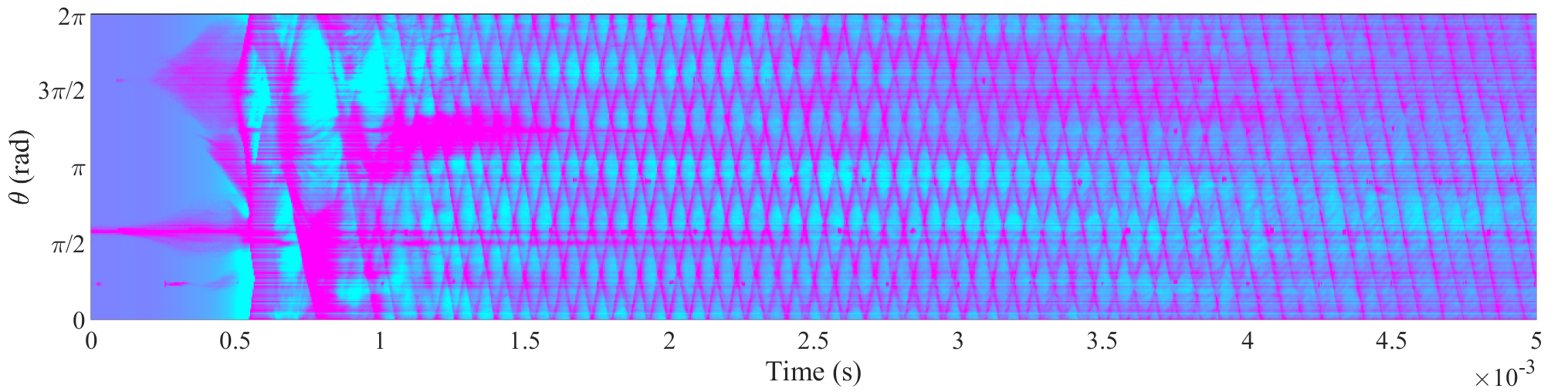}
	    \end{overpic}
        \caption{Representative ignition, deflagration-to-detonation transition, and mode-locking of detonation waves observed in an experiment. After ignition and an initial deflagration plume at 0.25 ms, a detonation wave forms at 0.5 ms that rapidly wraps around the annulus and consumes the majority of available propellant. After the injectors recover, a number of detonation waves nucleate and mode-lock into a regular pattern of three co-rotating waves and three counter-rotating waves. Eventually, one set of three waves weakens through asymmetric collisions and are eventually overrun. The final state is three co-rotating waves.}
		\label{fig:expStartup}
\end{figure*} 

Laboratory-grade RDEs are typically constructed with concentric cylinders to form a narrow annular combustion chamber (Fig. \ref{fig:intro}). At the head-end of the device, fuel and oxidizer are injected into the annulus (usually with separate fuel and oxidizer injectors) where they promptly mix to form a detonable medium. An ignition source (such as an automotive spark plug or torch) promotes an exothermal chemical reaction that rapidly releases heat into the fluid. Because of the confinement provided by the narrow annular gap, a rapid, local accumulation of energy occurs, forming steep gradients in pressure, density, and temperature in the fluid. The chemical kinetics associated with combustion are accelerated with temperature: these gradients therefore self-steepen, eventually forming shock waves coupled to regions of intense local energy release. These shock-reaction structures, or detonations, travel around the annulus of the engine the order of km/s. These structures persist so long as there is a sufficient supply of unburnt and well-mixed propellant to overcome competing dissipative effects, such as the rapid expansion of the flow downstream (a flameout) \citep{Wang2015,Yao2017,Jin2020} and heat transfer out of the combustion chamber \citep{Roy_2015,Theuerkauf2016,Goto2019}.

For successful engine operation, propellant injection and mixing must occur within the period of the detonation wave (or wavetrain, should multiple wave fronts be present) \citep{Wang2011,Nordeen2015,Duvall2018}. Similarly, the hot exhaust products need to be expelled before appreciable reintroduction of propellant can occur - otherwise, propellant injection may co-exist with hot products and promote parasitic deflagration (combustion not associated with a traveling wave - a loss mechanism \citep{Chacon2019,Chacon2019a}) within the combustion chamber. These associated physical processes of injection, mixing, exhaustion, and combustion all act on drastically different time and spatial scales to give the canonical RDE waveform (for example, if tracking pressure): a steep rise in pressure (detonation wave) that precedes a gradual decay back to an ambient condition \citep{Schwer2011,Schwer2019}. Combustion dominates the physics at the location of the detonation wave. However, once the propellant is consumed by the wave, exhaust and injection processes dominate the physics and the state decays back to an ambient condition on a time scale several orders of magnitude slower than that of the detonation wave. During this decay to an ambient condition, propellant regeneration is favored (although deflagration may occur along the contact surface between fresh propellant and hot combustion products \citep{Naples2013,Wang2016,Rankin2017a}). The modulation of injection and mixing is exacerbated by the feedback of the detonation waves into the reactant plenums \cite{KochPE,Koch2020}. A high-pressure detonation wave can temporarily block the injection process, as injection pressure can be lower than that of the detonation wave \citep{Schwer2012,Fotia2014,Driscoll2016,Sun2017,Zhou2018}.  The steady operation of the RDE is anomalous in that it is the separation of these associated time and spatial scales that enable stable operation, as concluded by several experimental and computational studies \cite{Hishida2009,Wu2014,Lu2014,Prakash2019}.

Several experimental programs have published detailed sweeps of geometry, injection schemes, and fueling conditions (for examples, see \citep{Dyer2012,Fotia2016,Fotia2017,Walters2019}) that establish several characteristic modes of operation. These include wave co-rotation with varying counts and direction, wave counter-propagation (both with equal and different number of co- and counter-propagating waves), periodic wave nucleation and extinction (`slapping' modes), and pulsating plane waves.

\subsection{Theoretical and Computational Considerations}
The exploration of RDE physics ultimately relies not only on experimentation, but also on detailed computational fluid dynamic simulations. Thus in a parallel line of work, computational fluid dynamics has been used extensively to diagnose the RDE flowfields with respect to similar parametric changes. These simulations vary from generic 2-D `unwrapped' domains (see \citep{Schwer2011,Schwer2019} for the standard premixed approach or \citep{Subramanian2020} for non-premixed) to full 3-D with detailed engine-specific geometries and injection schemes (see \citep{Gaillard2017,Sun2017,Sun2018,Lietz2018} as examples). When these simulations yield stable operation with periodic behavior, relevant thermodynamic metrics can be extracted, such as available mechanical work output and thermodynamic efficiency, from particle paths \citep{Zhou2012,Nordeen2014}.

Because the fastest physics (the detonation front) and the slowest physics (mixing and/or exhaustion) both need to be adequately resolved for proper system behavior \citep{Paxson2014,Cocks2016,Pal2019}, simulations need to be run for several - if not dozens or hundreds - of cycles. This ensures that the physics occuring on the slowest time scales can fully develop. The computational cost of simulations can quickly become prohibitive as one adds fidelity or model complexity. Consequentially, several research groups have developed reduced-order modeling approaches that adequately predict trends at a fraction of the computational cost. Such models exist for recreating the RDE canonical flowfield \citep{Fievisohn2017,Sousa2017a}, predicting thermodynamic trends \citep{Kaemming2017}, predicting application-based propulsive performance \citep{Mizener2017}, or reproducing the dynamics of the waves \citep{Humble2019,Koch2020,Koch2020a} with varying degrees of success. However, because of the multi-scale nature of the RDE and the intricate interactions of its fundamental physical processes, these modeling efforts are often constrained to geometry, propellant, or mode-specific operating regimes, with \textit{a priori} knowledge of wave topology or detonation structure. 

\subsection{Our contribution}
Our modeling efforts aim to establish a link between engineering performance metrics, such as mechanical work and thermal efficiency, to the nonlinear dynamics and wave topologies that are readily observed in experiments and widely reported in literature.  The work is motivated by our recent efforts in producing reduced order phenomenological models~\citep{Koch2020,Koch2020a} of the RDE {\em mode-locking} dynamics.  Specifically, we provide a first-principles, asymptotic derivation of the RDE dynamics starting from the Euler equations for  inviscid compressible flow constrained to one-dimension.  Importantly, the current work is able to capture the experimentally observed counter-propagating waves that were beyond description of our previous models~\citep{Koch2020,Koch2020a}.

%While the presented work is not an end-all resolution to the complexities of the RDE and its dynamics, we do believe it can provide valuable insight and trends that can inform future RDE design and analysis. 

Our work is motivated by direct observation of rotating detonation waves in an in-house designed and tested laboratory-scale gaseous methane-oxygen rotating detonation engine. This engine was fired in a test campaign that varied inlet and outlet boundary conditions parameterized by injector total area \citep{Koch2019,KochPE}. This test campaign resulted in wave structures that predominately featured counter-propagating waves equal in number (three clockwise and three counterclockwise, for example). However, for certain experiment operating points, co-rotating waves were observed, though their strengths (as inferred from wave velocity) were weak in comparison to the Chapman-Jouguet theoretical conditions. From high-speed imaging and pixel-intensity binning algorithms \cite{Bennewitz2018}, the kinematics of the combustion waves can be readily extracted and visualized, as in Figs. \ref{fig:intro} and \ref{fig:expStartup}.  These plots convey typical detonation wave behavior of the in-house experimental configuration. Fig.~\ref{fig:expStartup} details the space-time history for an experiment that ultimately resulted in three co-rotating waves. At ignition at about 0.25 ms, a deflagration plume grows. At the leading edges of the plume, detonations form at about 0.5 ms. These counter-propagating detonation waves quickly travel around the annulus at collide with each other. Because of the high strength of these initial waves, the injectors are temporarily blocked and no propellant exists to sustain the detonations after collision. The combustion is not halted, however. Once the injectors recover, a number of lesser-amplitude detonation waves form and self-organize into a regular pattern of three waves rotating clockwise and three waves rotating counter-clockwise. These waves persist until about 4 ms when one set of three waves weakens and is overrun by the opposing set of waves.  As shown, our derived model is capable of characterizing the experimentally observed dynamics of the RDE, including its rich combustion wave interactions and both co- and counter-propagating wave dynamics.

%In Section \ref{sec:background}, we begin with an overview of the rotating detonation engine hardware and its operation as well as a brief review of current trends reported in literature. Included in Section \ref{sec:background} is an example kinematic trace of waves formed during an in-house hot-firing of a laboratory-grade RDE.
%The paper is outlined as follows:
%The construction of a model for the nonlinear dynamics is given in Section \ref{sec:model}, followed by results of numerical experimentation of the model in Section \ref{sec:results}. We end the paper with a discussion of the RDE, its dynamics, and the presented model in Section \ref{sec:discussion}.

\section{Lumped-Volume Double-Choke Combustor Model} \label{sec:model}
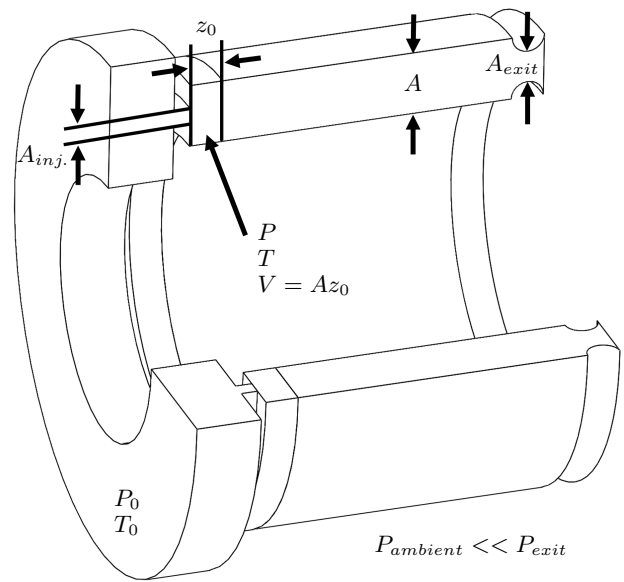
\begin{figure}[t]
        \centering
        \begin{overpic}[width=1\linewidth]{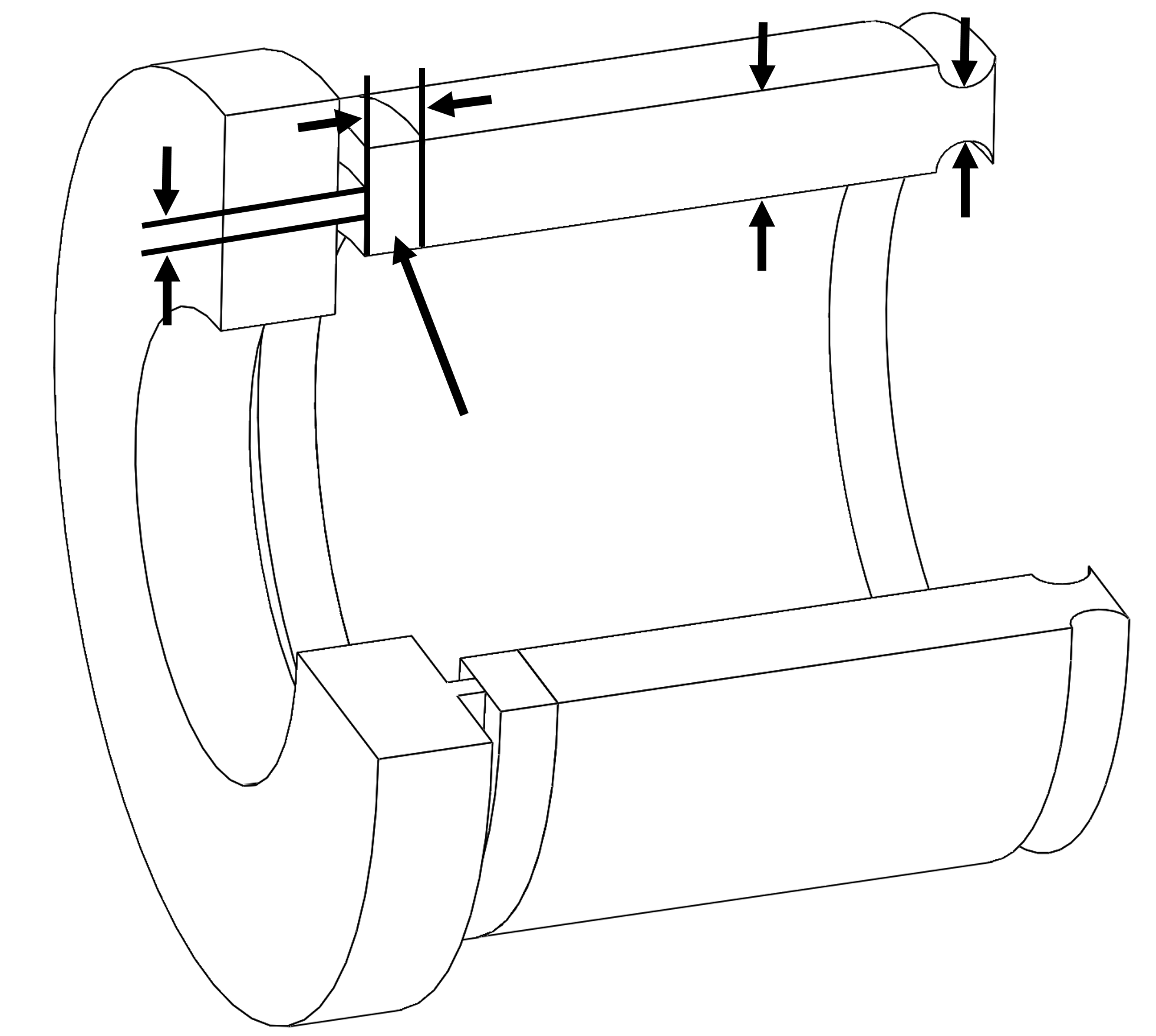}
        \put(5,65){$A_{inj.}$}        
        \put(64.5,76){$A$}  
        \put(77,79){$A_{exit}$}  
        \put(20,12){$P_{0}$}
        \put(20,8){$T_{0}$}  
        \put(60,5){$P_{ambient} << P_{exit}$}  
        
        \put(42,53){$P$}
        \put(42,49){$T$}          
        \put(42,45){$V = A z_0$}         
        
        \put(32.5,85){$z_0$}          

	    \end{overpic}
        \caption{Schematic for the derivation of the zero-dimensional double-choked reactor model. Consider a slice through the annular combustor, as shown by the exposed face of the displayed section cut of a notional RDE. Propellant is injected from a pseudo-infinite reservoir with pressure $P_0$ and temperature $T_0$ through an injection area $A_{inj.}$. The propellant rapidly mixes and combusts in a thin combustion layer (with depth $z_0$ and cross-sectional area $A$) attached to the front endwall of the engine. The pressure $P$ and temperature $T$ associated with this thin layer dynamically responds to injector-combustor coupling and exhaust processes. The ambient backpressure $P_{ambient}$ is assumed to be sufficiently low to always induce exhaust flow thermal choking. The exhaust products expand isentropically from the state of the thin combustion layer to the sonic condition at the exit of the combustor.}
		\label{fig:zeroD}
\end{figure}

The goal of the modeling effort is to create a simplified framework from which the nonlinear dynamics associated with rotating detonation waves can be reproduced and related to engineering performance metrics. Our approach combines aspects of the rotating detonation analog \cite{Koch2020}, which imposed a global input/output energy balance subject to a shock-forming medium, with the one-dimensional (1-D) Euler equations for inviscid compressible flow. The Euler equations are used to model the fluid flow along the circumference of the annulus of an RDE. At each spatial point along this domain, a lumped-volume combustor model is used to impose an input/output energy and mass balance and to model propellant mixing. The two models are coupled through source terms for the 1-D Euler equations, allowing the model domain to remain exclusively on the 1-D periodic line. 

We begin the construction of the model by examining a slice of a generic RDE annular combustion chamber along its axis, as depicted in Fig. \ref{fig:zeroD}. From left to right, the major features of this cross-sectional slice include the notional injector that is fed by a pseudo-infinite reservoir of gaseous propellant at pressure $P_0$ and temperature $T_0$ with injection area of $A_{inj.}$. The injector is nominally choked. The combustion chamber as depicted is divided into two sections: the thin heat release zone and the large expansion region. The combustion chamber has cross-sectional area $A$. The expansion region terminates in a notional geometric throat with area $A_{exit}$. The exhaust gases are expelled to an ambient condition which is assumed to be a vacuum. Consequentially, the Mach number at the exit of the combustor is one for all spatial locations and for all time.

To further simplify the model domain, we assume that the driving physics occur within the thin layer attached to the front endwall of the combustor. The depth of this layer is on the same order as a small length scale, $z_0 = V/A \approx O(\epsilon)$. Furthermore, we enforce a zero axial velocity boundary condition at the base of the layer. Although an axial pressure gradient exists (as seen in experimental studies \citep{KochPE}), the change in axial velocity over this short length is of order $O(\epsilon)$ and the contributions to momentum and kinetic energy are of orders $O(\epsilon)$ and $O(\epsilon^2)$, respectively. Thus, we neglect axial momentum at this location as well as contributions of axial velocity to internal energy. Note that is assumption breaks down (i) as one moves further downstream from the heat release zone, where significant energy conversion to kinetic energy occurs and comprises a significant portion of internal energy, and (ii) for injection schemes with appreciable axial momentum. Nevertheless, we adopt this fundamental assumption to proceed with model simplification. The region of interest is now collapsed to this thin layer at the front endwall of the engine. We proceed by treating this layer at any arbitrary azimuthal location as a lumped volume. 

We write the conservation of mass, energy, and the combustion progress variable for this lumped volume domain as a set of coupled ordinary differential equations (as opposed to partial differential equations with spatial dependence) for the conserved quantities $\mathbf{Q}$:

\begin{equation} \label{eq:ode}
\frac{\partial \mathbf{Q}}{\partial t} = \frac{d \mathbf{Q}}{d t} = \mathbf{S}
\end{equation}

\begin{equation} \label{eq:variables}
\mathbf{Q} = \begin{bmatrix}
\rho \\ E \\ \rho \lambda
\end{bmatrix}
\end{equation}
\begin{equation} \label{eq:sourceDim}
\mathbf{S} = \begin{bmatrix}
\frac{1}{V}\left({m}^{+} - {m}^{-}\right) \\ \frac{1}{V}\left({m}^{+}e_{0} - {m}^{-}e\right) + \omega q \\ \omega - \rho \lambda \beta + \frac{\lambda}{V}\left({m}^{+} - {m}^{-}\right)
\end{bmatrix},
\end{equation}
where $V$ is the volume of the domain, $m^+$ and $m^-$ are the mass flow rates into and out of the domain, $e$ is specific internal energy, $E$ is total energy, $\omega$ is the chemical reaction rate, $q$ is the volumetric heat release associated with the propellant, $\rho$ is density, $\lambda$ is the combustion progress variable, and $\beta$ is the propellant injection and mixing model. In Eqs. \ref{eq:ode}-\ref{eq:sourceDim}, quantities with subscript ${\left(\cdot\right)}_0$ denote injection reservoir quantities. 

Density is related to pressure and temperature through the ideal gas law:
\begin{equation}
P = \rho R T ,
\end{equation}
where $R = c_p - c_v$ is the specific gas constant for the fluid, $c_v$ is the specific heat at constant volume, and $c_p$ is the specific heat at constant pressure. Furthermore, the assumption of a calorically perfect gas is made, enabling the specific internal energy to be written as $e = c_v T$, or alternatively as:
\begin{equation} \label{eq:internalEnergy}
e = \frac{P}{\rho \left(\gamma - 1\right)},
\end{equation}
where  $\gamma = c_p/c_v$ is the nondimensional ratio of specific heats. 

The source term $S$ reflects the input and output pathways for mass and energy. Mass flow occurs at the injection plane (${m}^+$) and at the exit plane of the domain (${m}^{-}$) with corresponding areas $A_{injection}$ and $A_{exit}$. Similarly, the flow of energy occurs along these pathways with the addition of a source for head addition through chemical reactions. Lastly, the dynamics of the combustion progress variable $\lambda$ follow the combustion and regeneration balance ($\omega$ versus $\beta$) subject to the fluctuations of mass inside of the domain (chain rule). 

\begin{figure}[t]
        \centering
        \begin{overpic}[width=1\linewidth]{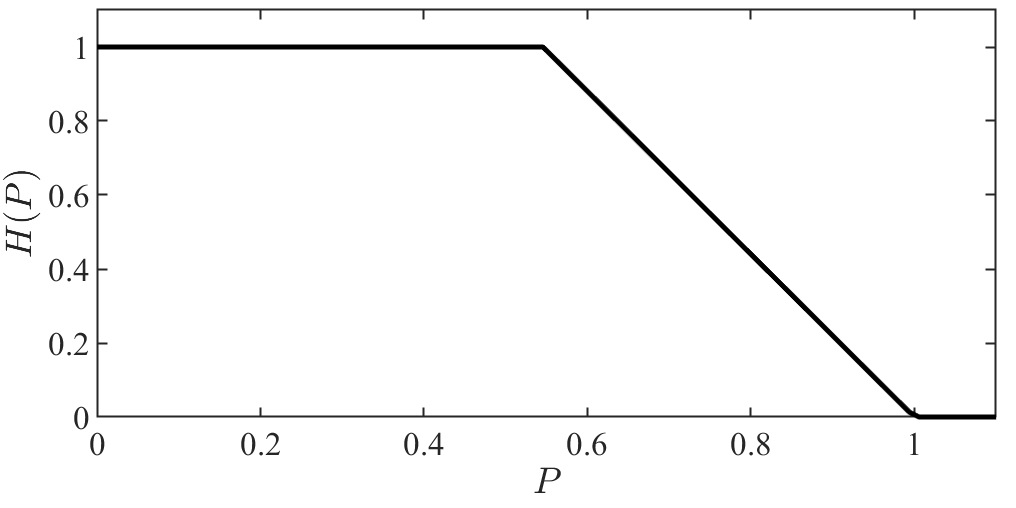}
	    \end{overpic}
        \caption{Activation function $H(P)$ for $\gamma=1.3$. The `knee' in the curve occurs at the value of $P$ that induces choked flow.}
		\label{fig:activation}
\end{figure}

\begin{figure}[t]
        \centering
        \begin{overpic}[width=1\linewidth]{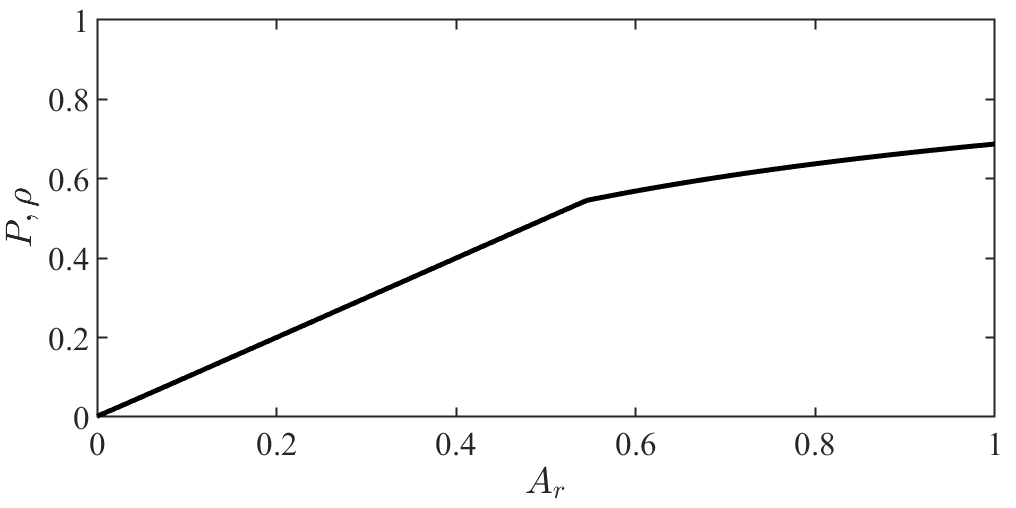}
	    \end{overpic}
        \caption{Steady-state operating levels of pressure and density for the lumped-volume model as the area ratio of the model is varied. Note that because of the formulation of the lumped-volume method, $T=1$ in the steady state (when no chemical reactions occur).}
		\label{fig:fixedPts}
\end{figure}

\subsection{Mass Fluxes} \label{sec:rate1}

The mass flow rate into the domain is assumed to be governed by choked-flow relationships. However, known is that the chamber pressure can reach or exceed the injection manifold pressures. For the present model, we employ a simplified treatment of injection modulation: If the pressure in the domain is sufficiently low, the mass flow rate into the domain is modeled by choked injection, or a constant rate. If the chamber pressure is equal or greater than of that of injection, the injector is assumed to be blocked and no mass is introduced into the domain. In making this approximation, we neglect any injection losses that may occur through shock formation or friction. We reiterate our previous assumption that injection kinematics do not affect the evolution of momentum in the domain. 

The relationship for the rate of mass flow into the domain is given by:
\begin{equation}
{m}^+ = \rho^{*} v^{*} A_{injection} ,
\end{equation}
where the superscript ${\left(\cdot\right)}^*$ denotes sonic quantities. $v^*$ is the sonic velocity at the injection area minimum and $A_{injection}$ is the magnitude of the injection area minimum. Because the flow is assumed to be choked at the throat of the injector, the relationship can be rewritten with the introduction of isentropic flow relationships and using the propellant manifold quantities:
$$
m^+ = A_{injection} \sqrt{P_{0} \rho_{0}} \sqrt{\gamma}\left(\frac{2}{\gamma + 1}\right)^{\frac{\gamma + 1}{2\left(\gamma - 1\right)}},
$$
\begin{equation} \label{eq:massFlow}
m^+= A_{injection} \sqrt{P_{0} \rho_{0}} I_c ,
\end{equation}
where $I_c$ is defined by the constant function of $\gamma$.

This relationship is modulated by applying an activation function as a switch and ramp to regulate mass flow rate in the presence of strong chamber pressure fluctuations:
\begin{equation} \label{eq:activation}
H(P) = \mathcal{H}\left(1 - \frac{P}{P_{0}}\right)\left( 1 - \mathcal{H}\left( \frac{P}{P_{0}} - r \right)  \frac{P/P_{0} - r}{1-r}\right)
\end{equation}
\begin{equation}
r = {\left(1 + \left(\frac{\gamma - 1}{2}\right)\right)}^{\frac{-\gamma}{\gamma - 1}} ,
\end{equation}
where $\mathcal{H(\cdot)}$ is the Heaviside step function. The behavior of this activation function is shown in Fig. \ref{fig:activation}. The knee in the curve of the activation function occurs at $r$, the magnitude of the pressure ratio required to induce choking. For pressure ratios with magnitudes below $r$, the mass flux is `choked' and unaffected by the conditions in the domain. If the pressure ratio falls between one and $r$, it obeys a linear ramp to zero mass flux at $P/P_{inj} = 1$. Beyond $P/P_{inj} = 1$, $H$ is zero: injection and mixing are halted.

The mass flow rate is therefore:
\begin{equation}\label{eq:injMod}
m^+ = A_{injection} I_c H. 
\end{equation}

Note that this treatment neglects all injection kinematics and associated losses (such as shock structures at or near the exits of the injectors and frictional effects, for example). 

Recall that the ambient backpressure is a vacuum. Assumed is that the exit of the combustor is choked (Mach number of one) at all spatial locations for all time. Treating this as a boundary condition, we relate the exit conditions to those of the lumped volume through the same isentropic flow relations used in Eq.\ref{eq:massFlow}.  Mass flow out of the \textit{combustor} is given by:
\begin{equation}
m^- = \rho^* v^* A_{exit} ,
\end{equation}
where $v^*$ is the velocity at the exit of the domain and $A_{exit}$ is the cross-sectional domain exit area. This relationship is rewritten by invoking the choked exit boundary condition:
\begin{equation}
m^- = A_{exit} \sqrt{P \rho} I_c .
\end{equation}
Furthermore, we assume that there is no energy or mass accumulation in the combustor. Thus, the mass flow out of the combustor is equal to the mass flow out of the lumped volume.  Because the flow is assumed to expand isentropically from the lumped volume to the choked combustor exit, internal energy remains constant through this process. The energy flux out of the combustor is therefore taken to be the energy flux out of the lumped volume. This outlet energy flux is given by:

\begin{equation}
    m^- e = \frac{P}{\rho(\gamma-1)} A_{exit} \sqrt{P \rho} I_c .
\end{equation}

\subsection{Combustion}
Combustion is prescribed by a single-step kinetic model that is Arrhenius in type. The rate of reaction is given as:
\begin{equation}
K(T) = k \exp\left(\frac{-E_a}{RT}\right) ,
\end{equation}
where $k$ is the reaction rate constant, $E_{a}$ is the activation energy, and $T$ is the temperature in the domain. The rate of energy input into the domain from chemical reactions is therefore:
\begin{equation}
\omega q= K\left(T\right) \rho \left(1-\lambda\right) q,
\end{equation}
where $q$ is the volumetric heat release associated with the propellant. The depletion of reactant corresponds to $\lambda$ approaching $\lambda = 1$.

\subsection{Injection and Mixing} \label{sec:rate3}

The regeneration of $\lambda$ (pushing $\lambda$ toward zero) is influenced by the separate processes of injection and mixing. In the simplest model, a time scale for mixing ($s$) is imposed, but modulated with the activation function of Eq. \ref{eq:injMod}:

\begin{equation}
\beta = \frac{H}{s} .
\end{equation}
Thus, should there be no mass flow into the domain, there is no regeneration of $\lambda$. Similarly, as $H$ is activated, the rate of $\lambda$ regeneration approaches $1/s$. In general, $s$ may depend on a wide array of parameters, such as mass flux, equivalence ratio, injector geometry, etc.

This formulation is unique in that the inlet mass flow and rate constant for mixing are only loosely coupled through the activation function $H(P)$: a wide variety of mixing and injection time scales can be evaluated without prescribing detailed injection models. Similarly, by modulating the regeneration of $\lambda$ with $H$, a \textit{refractory period} (a region of space and time behind a detonation wave where there exists no detonable mixture) is imposed. This replicates the refractory periods seen in the physics of real RDEs as well as in the inlet boundary conditions of detailed 2-D and 3-D RDE CFD simulations. 

\subsection{Nondimensionalization} \label{sec:nondim}
The aforementioned rates and the coupled differential equation (Eq. \ref{eq:ode}) are non-dimensionalized based on injection manifold quantities ($P_{0}$ and $\rho_{0}$). Additionally, a length scale is defined by $z_0 = V/A$, a characteristic velocity by $u_{0} = \sqrt{ P_{0}/\rho_{0}}$, and a time scale by $t_{0} = z_0/u_{0}$. Introducing $\widetilde{\left(\cdot\right)}$ as the notation for non-dimensional quantities, we write:
$$
\widetilde{E} = \frac{E}{P_{0}} ~~~ \widetilde{\rho} = \frac{\rho}{\rho_{0}} ~~~ \widetilde{P} = \frac{P}{P_{0}}
$$
$$
\widetilde{t} = \frac{t}{t_{0}}~~~\widetilde{u} = \frac{u}{u_{0}} ~~~ \widetilde{\omega} = \omega \frac{t_{0}}{\rho_{0}}
$$
\begin{equation}
\widetilde{E_a} = E_a \frac{\rho_{0}}{P_{0}}~~~\widetilde{q} = q \frac{\rho_{0}}{P_{0}} ~~~ \widetilde{T} = T \frac{R \rho_{0}}{P_{0}} .
\end{equation}

We continue by substituting these relationships into the components of the coupled ODE of Eq. \ref{eq:ode}:
$$
\frac{d}{d\widetilde{t}}\begin{bmatrix}
\widetilde{\rho} \\ \widetilde{E} \\ \widetilde{\rho} \lambda
\end{bmatrix}  = \mathbf{\widetilde{S}}
$$
\begin{equation} \label{eq:source}
\mathbf{\widetilde{S}} = 
\begin{bmatrix}
\alpha\left( A^+ H(\widetilde{P}) - A^- \sqrt{\widetilde{P}\widetilde{\rho}}\right)
\\
\frac{\alpha}{\gamma - 1}\left(A^+ H(\widetilde{P}) - \widetilde{T} A^- \sqrt{\widetilde{P}\widetilde{\rho}}\right) + \widetilde{\omega} \widetilde{q}
\\
\widetilde{\omega} + \widetilde{\rho}\widetilde{\beta} H(\widetilde{P}) \lambda + \alpha\left( A^+ H(\widetilde{P}) - A^-\sqrt{\widetilde{P}\widetilde{\rho}}\right)\lambda
\end{bmatrix} ,
\end{equation}

where the dimensionless groups $\alpha$ and $\widetilde{\beta}$ are defined as:

\begin{equation}
\alpha = \frac{\sqrt{p_0 \rho_0}I_c}{\rho_{0} u_{0}}
\end{equation}

\begin{equation}
\widetilde{\beta} =  \frac{t_{0}}{\widetilde{s}} .
\end{equation}

The area profile through the combustor is provided by relating an overall area ratio, $A_r$, to a blockage ratio, $c$:

\begin{equation}
A_r = \frac{A_{injection}}{A_{exit}},
\end{equation}

\begin{equation}
    \frac{A_{exit}}{A_{chamber}} = A^- = \left( 1 - c \right)
\end{equation}

\begin{equation}
    \frac{A_{injection}}{A_{chamber}} = A^+ = A_r A^-.
\end{equation}

Setting $c=0$ recovers a straight annular duct profile. A 10\% blockage in exit area corresponds to $c=0.1$, etc. The reaction rate $\widetilde{\omega}$ is recast with a characteristic reaction rate and temperature, $k = \widetilde{k}k_{0}$ and $\widetilde{T}_{ign.}$:

\begin{equation}
\widetilde{\omega} = Da \widetilde{\rho} (1-\lambda) \widetilde{K}(\widetilde{T})
\end{equation}

\begin{equation}
\widetilde{K}(\widetilde{T}) = \exp\left( -\widetilde{E_a} \left(\frac{1}{\widetilde{T}} - \frac{1}{\widetilde{T}_{ign.}} \right) \right)
\end{equation}

\begin{equation}
Da = t_{0} k_{0}
\end{equation}

\begin{equation}
k_{0} = k \exp\left(\frac{-E_a}{RT_{ign.}}\right)
\end{equation}

\begin{equation}
\widetilde{T}_{ign.} = \frac{T_{ign.} R \rho_{0}}{P_{0}} .
\end{equation}

The dimensionless group $\alpha$ is interpreted as the ratio of mass flow for the reference condition through the injector area versus that of the annular area. The dimensionless group $\widetilde{\beta}$ is the ratio of the flow convective time scale to the time scale for propellant mixing. The Damk\"ohler number, $Da$, relates the chemical time scale to the fluid convective time scale. Thus, the presented model explicitly relates the convective time scale of the fluid, the time scales of the mixing and injection processes, and the time scale of the chemical kinetics in a compact framework that is readily analyzed. 

From this point forward, and in all figures, all quantities are nondimensional. As such, we drop the $\widetilde{\left(\cdot\right)}$ notation for readability. Lastly, because our goal is not to reproduce a particular experimental observation, experimental apparatus, or behavior of a specific chemical propellant, we leave the model equations parameterized by these nondimensional groups and explicitly vary them. This allows for numerical experimentation through wide ranges of physical regimes without the constraint of device-specific particularities.  

\subsection{Dynamic Behavior}
The stand-alone zero-dimensional combustor model is useful for several reasons. First, the fixed points of the cold (no reactions) system give the baseline operating condition of the combustor; that is, the steady-state pressure and density of the medium through which the detonation waves travel. This operating condition is the result of a balance of inlet and outlet mass flows. Second, the model can adequately relate the the scales of injection, mixing, and exhaustion to give a dynamic response to an impulse (an internal explosion or detonation, for example). Pressure settling times, propellant recovery times, and chamber mass fluctuations can be readily evaluated with respect to model parameters. 

\subsubsection{Cold Flow and Impulse Response}
\begin{figure}[t]
        \centering
        \begin{overpic}[width=1\linewidth]{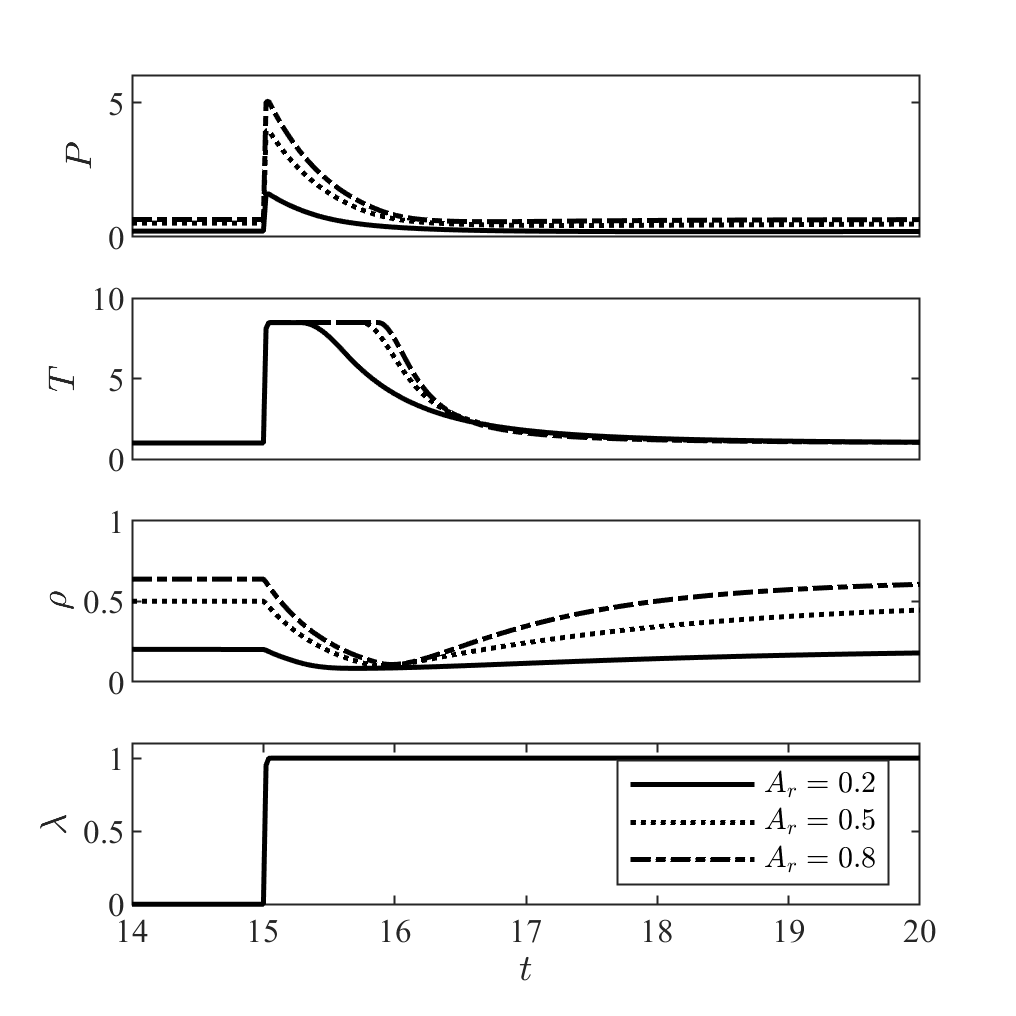}
        \put(13,95){(a)}
        \put(13,72.5){(b)}    
        \put(13,50.5){(c)}    
        \put(13,29){(d)}
	    \end{overpic}
        \caption{The refractory period imposed by the increase in chamber pressure is heavily influenced by the area ratio of the system. $E_a = 10$ for all simulations.}
		\label{fig:AR_variation}
\end{figure}
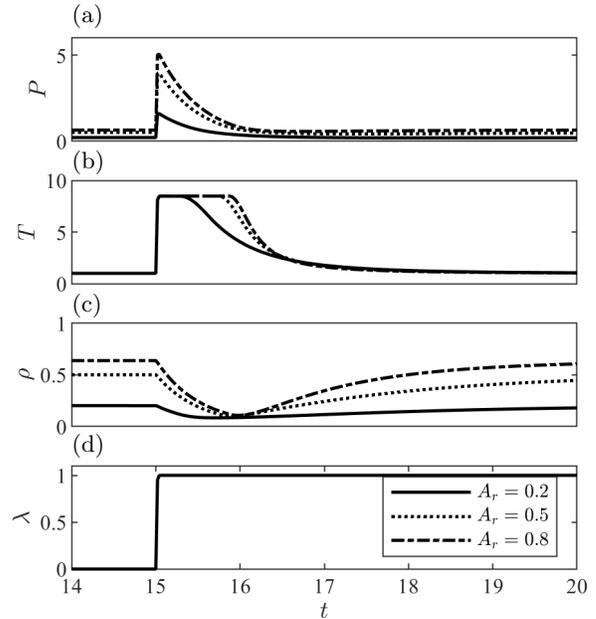

Mass flow through the combustor is predominately controlled by the injection manifold pressure and the injection area subject to the rapid exhaustion to an ambient condition. The balance of input and output mass flows set the operating level of the model. The set of steady-state operating points through injection-to-annulus area ratio is shown in Fig. \ref{fig:fixedPts}. These points are for the system without chemical reactions. The transient response of the lumped volume can be investigated by inducing an impulsive combustion event to the steady state of the model. Figure \ref{fig:AR_variation} shows representative time histories of such models. The combustion impulse is applied at $t = 15$ by setting the reference pre-exponential factor to an artificially high value. For this set of simulations, the regeneration of propellant is excluded to isolate the impulse response of the model. Parameters used for these simulations are listed in Table \ref{tab:params0D}. Figure \ref{fig:AR_variation} shows the behavior of the system for variations in area ratio alone. The peak pressure observed scales with area ratio, with larger $A_r$ exhibiting a larger peak pressure. Higher $A_r$ exhibits the greatest influence of injector blocking, as shown in Fig. \ref{fig:AR_variation}b. Injector blocking induces a \textit{refractory period} behind the combustion impulse which is seen immediately after $t=15$ until mass flow is reintroduced into the system. Refractoriness occurs when the activation function $H(P)=0$: only terms in $\mathbf{S}$ containing $\sqrt{{P}{\rho}}$ are active. Thus, during this period there is a decrease of pressure and density of the same proportion. Because $T = P/\rho = constant$, the associated expansion process is \textit{isothermal} but not \textit{isentropic}. 

Note that the inlet and outlet mass and energy flow pathways follow choked flow relationships derived from relating reservoir quantities to a sonic condition via isentropics. However, an implicit assumption was made that the temperature during these blowdown processes are \textit{isothermal} rather than \textit{adiabatic}. This assumption was made for several reasons. First, under this assumption the mass inside the lumped volume can be decoupled from temperature and energy dynamics, obeying simple time constants ($\sqrt{P_0 \rho_0} I_c$ and $\sqrt{P \rho} I_c$ ). Consequentially, the functional forms of the terms in $\mathbf{S}$ are able to be simplified with common terms such as the dimensionless group $\alpha$. Lastly, and most importantly, this assumption allows the feedback between combustion (which is solely a function of temperature), injection, and exhaustion to be isolated and compared equivalently across all geometric changes ($A^+$,$A^-$,$A_r$, and $z_0$). Although the time constants for injection and exhaustion may change, the chemical kinetic model is unaffected by changes in the area ratios. The major system-wide consequence of this assumption is entropy generation during injection and exhaustion.

\begin{table}
	\caption{0-D Model Parameters}
	\label{tab:params0D}
	\centering
	\begin{tabular}{ccccccccccc}
	\hline
	$\gamma$ &$q$& $Da$ & $\beta$ & $T_{ign.}$ & $c$ \\
	
	1.3 & 25 & 10 & 0.1 & 3 & 0 \\

	\hline
	\end{tabular}
\end{table}

\subsubsection{Propellant Regeneration and Parasitic Deflagration} \label{sec:0D_combustion}
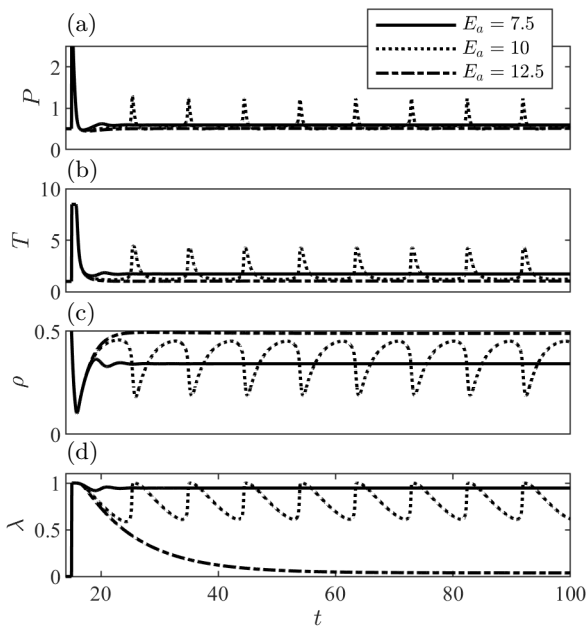
\begin{figure}[t]
        \centering
        \begin{overpic}[width=1\linewidth]{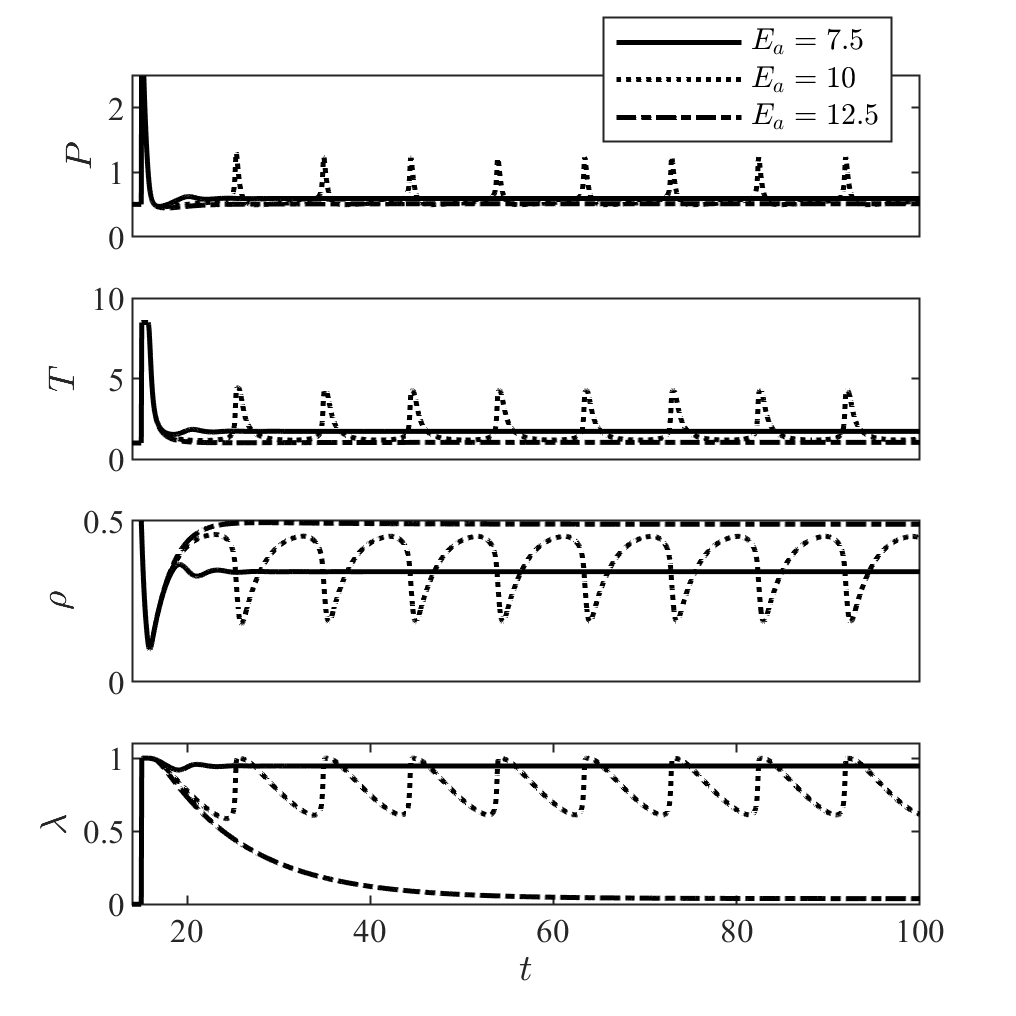}
        \put(13,95){(a)}
        \put(13,72.5){(b)}    
        \put(13,50.5){(c)}    
        \put(13,29){(d)}
	    \end{overpic}
        \caption{Variation of activation energy. The amount and rate of combustion on the decaying side of the impulse can lead to steady deflagration ($E_a = 7.5$), limit cycling ($E_a = 10$), or the return to the original state of the system ($E_a=12.5$). $A_r=0.5$ for all simulations. Note that qualitatively similar behavior can be obtained by alternatively varying $Da$, $\beta$, or $c$.}
		\label{fig:EA_variation}
\end{figure}

Propellant flows into the combustion chamber when the chamber pressure is sufficiently low, corresponding to the activation function of Eq. \ref{eq:activation}. The interaction between the state of the domain, the chemical kinetic model, and propellant refractory period can result in one of three scenarios: (i) the refractory period is long enough such that the chemical kinetics are not activated upon injection (the combustor temperature becomes low) and $\lambda$ approaches 0 according to the regeneration time constant and mass flux, (ii) the refractory period is short, causing propellant to be introduced while combustor conditions allow for activated kinetics, meaning that the rate of combustion can offset the rate of injection and mixing so that $0 < \lambda < 1$, or (iii) the combined processes of injection and mixing oscillate in-phase with combustion and exhaustion processes (limit cycling). All three scenarios have been experimentally observed \citep{KochPE}. While a complete survey of parameters and bifurcations within these regimes is not within the scope of this paper, we do present example time histories of each of these regimes. In Fig. \ref{fig:EA_variation}, three time histories are shown depicting the same model combustor configuration as in Fig. \ref{fig:AR_variation}, though fixing the area ratio at 0.5 and allowing propellant to regenerate after the combustion impulse. Three activation energies are displayed, each leading to one of the three aforementioned scenarios. Each time history trace possesses the same refractory period, as the area ratio and volumetric heat release is fixed. Additionally, the dimensionless group $\beta$ is held constant at 0.1. For the lowest activation energy, $E_a=7.5$, chemical reactions continue immediately upon reintroduction of propellant after the end of the refractory period. After oscillations in the state of the model decay, the system approaches a steady state where the rate of propellant regeneration matches the rate of heat release in the domain, subject to the mass and energy flux balances of the model. For $E_a=10$, combustion is slower upon the reintroduction of propellant, though through time the reactions accelerate and eventually dominate the dynamics of the system, causing a sharp rise in temperature and pressure as the propellant is depleted. This process repeats periodically. Lastly, for the highest presented activation energy of $E_a=12.5$, the system returns to its original state after the combustion impulse. The heat release rate is small compared to the rate of energy flux out of the domain. Similarly, the rate of propellant depletion is small relative to the regeneration rate. The combustion that occurs on the decaying side of the impulse is termed \textit{parasitic deflagration} \citep{Chacon2019,Chacon2019a}. We note that qualitatively similar transitions between these regimes can occur by instead exclusively varying $\beta$, $Da$, or $c$. 

\section{Spatially Extended  Model} \label{sec:results}
We extend the zero-dimensional model to the one-dimensional (1-D) line by applying the source terms of the zero-dimensional model to the 1-D reactive Euler equations (after consistent nondimensionalization):

\begin{equation} \label{eq:pde}
\frac{\partial}{\partial t}\begin{bmatrix}
\rho \\ \rho u \\ E \\ \rho \lambda
\end{bmatrix} 
+
\frac{\partial}{\partial x}\begin{bmatrix}
\rho u \\ \rho u^2 + P \\ u\left(E + P\right) \\ \rho u \lambda
\end{bmatrix}
= \mathbf{S} .
\end{equation}

$\mathbf{S}$ is the source term vector of Eq. \ref{eq:ode} extended to include the momentum equation:

\begin{equation}
\mathbf{{S}} = 
\begin{bmatrix}
\alpha\left( A^+ H({P}) - A^- \sqrt{{P}{\rho}}\right)
\\
0
\\
\frac{\alpha}{\gamma - 1}\left(A^+ H({P}) - {T} A^- \sqrt{{P}{\rho}}\right) + {\omega} {q}
\\
\widetilde{\omega} + {\rho}{\beta} H({P}) \lambda + \alpha\left( A^+ H({P}) - A^-\sqrt{{P}{\rho}}\right)\lambda
\end{bmatrix} ,
\end{equation}

Because of the lumped-volume approach, after a fluid particle exits the modeled thin layer it is no longer in communication with fluid particles of different azimuths. Therefore, to enforce system-wide conservation of momentum in the azimuthal direction, we prescribe it explicitly in Eq. \ref{eq:pde} where there is communication via spatial derivatives. The total energy of the fluid is extended to include kinetic energy associated with circumferential velocity:
\begin{equation}
E = \frac{P}{\gamma-1} + \frac{1}{2}\rho {u}^2 .
\end{equation}
To be consistent with the treatment of conservation of azimuthal momentum, the form for outlet energy flux in $\mathbf{S}$ remains unchanged (although internal energy now has a kinetic energy component). Azimuthal kinetic energy is not penalized; only the static temperature and pressure at each spatial location.

For periodic boundaries, this spatially-extented model will find the distribution of energy release subject to the behavior of the zero-dimensional combustor model, i.e., dynamic injection and exhaustion responses. Desired are model states where the distribution of energy results in the formation of stable, coherent combustion wave fronts - rotating detonation waves. 

In this section, three separate sets of numerical experiments are presented. First, we verify the ability of the numerics to capture the Zeldovich-von Neumann-D\"oring structure of a traveling 1-D detonation wave and perform a mesh convergence study. This is performed in a 1-D model detonation tube. Second, evaluated are detonation profiles of the full model equation with input/output flow in the same detonation tube setting. Third, exploratory numerical sweeps are conducted on the periodic line with the goal of exhibiting model properties and pattern formation within the system. 

\subsection{Numerical Validation and Convergence} \label{sec:validation}

\begin{figure}[t]
        \centering
        \begin{overpic}[width=1\linewidth]{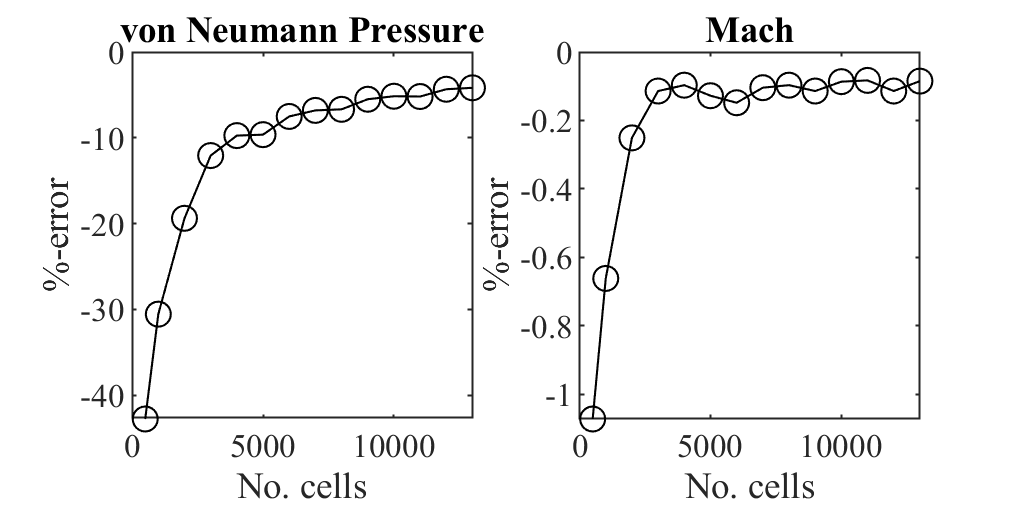}
        \put(13,41.5){(a)}   
        \put(57,41){(b)}
	    \end{overpic}
        \caption{In (a), the error in peak pressure is shown for a set of different mesh resolutions. In (b), the error in wave Mach number is shown. The convergence threshold is set to maintaining  less than a 5\% error in peak pressure. The corresponding mesh resolution is $\Delta x = 5 \cdot 10^{-3}$. }
		\label{fig:convergence}
\end{figure}

\begin{figure}[t]
        \centering
        \begin{overpic}[width=1\linewidth]{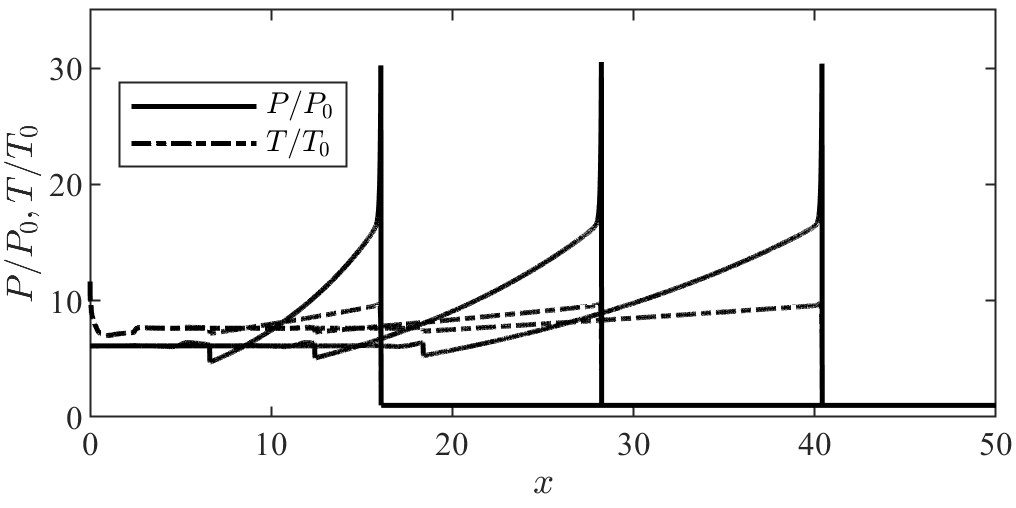}
	    \end{overpic}
        \caption{Detonation profiles for validation test case at times $t=4$, $6$, and $8$.}
		\label{fig:validation}
\end{figure}

A numerical detonation tube is simulated to provide confidence in the numerical treatment of the model. Numerical computations are performed with PyClaw \citep{Ketcheson2012} with a time-splitting method for integrating the stiff source terms. A Harten-Lax-van Leer (HLLC) Riemann solver is used to evaluate fluxes. The source term integrator is a two-stage, second-order Runge-Kutta scheme. The numerical detonation tube experiment consists of the 1-D domain that is closed on one end (zero-gradient wall at $x=0$) and open on the opposite end (zero-order extrapolation at $x=L$). The detonation tube is initialized with $P(x,t=0) = 0.5$, $T(x,t=0)=1.0$, and $\lambda(x,t=0) = 0$. The chemical kinetic model is unchanged from that of Section \ref{sec:0D_combustion} with $q=25$, $E_a=10$, $Da=10$, and $T_{ign.}=3.0$. To start reactions, the first computational cell in the domain is initialized to a high temperature ($T = 5.0$). The length of the detonation tube for this set of numerical experiments is $L=50$. The number of cells along this dimension is varied from 500 to 13000. To quantitatively evaluate the resultant detonation profiles, the peak pressure and wave Mach number are compared with the respective theoretical values from Chapman-Jouguet and Zeldovich-von Neumann-D\"oring theories. When nondimensionalized as in Section \ref{sec:nondim}, the wave Mach number is given by:

\begin{equation}
M_{{CJ}^+} = \sqrt{1 + q\frac{\gamma^2-1}{2\gamma}} + \sqrt{q \frac{\gamma^2 - 1}{2\gamma}} ,
\end{equation}
which evaluates to $M_{{CJ}^+}=5.339$ for $q=25$ and $\gamma = 1.3$. The von Neumann peak pressure is obtained through normal shock relationships, yielding a shock pressure ratio of 32.09 for this Mach number. Figure \ref{fig:convergence} profiles the error in the developed peak pressures and detonation Mach number as a function of the mesh resolution. Resolving the von Neumann spike drives the mesh resolution requirement - the numerics capture wave speeds remarkably well, even with the coarsest meshes and the low-order scheme. Therefore, our convergence criteria is arbitrarily set to obtaining less than a 5\% error in peak pressure. This threshold is reached at a grid resolution of $\Delta x = 50/10000 = 5 \cdot 10^{-3}$. All numerical results in this paper are presented with mesh resolutions corresponding to this spacing. Figure \ref{fig:validation} displays the resultant detonation profiles at times $t = 4$, $6$, and $8$.

\subsection{Model Detonation Tube Profiles}
\begin{figure}[t]
        \centering
        \begin{overpic}[width=1\linewidth]{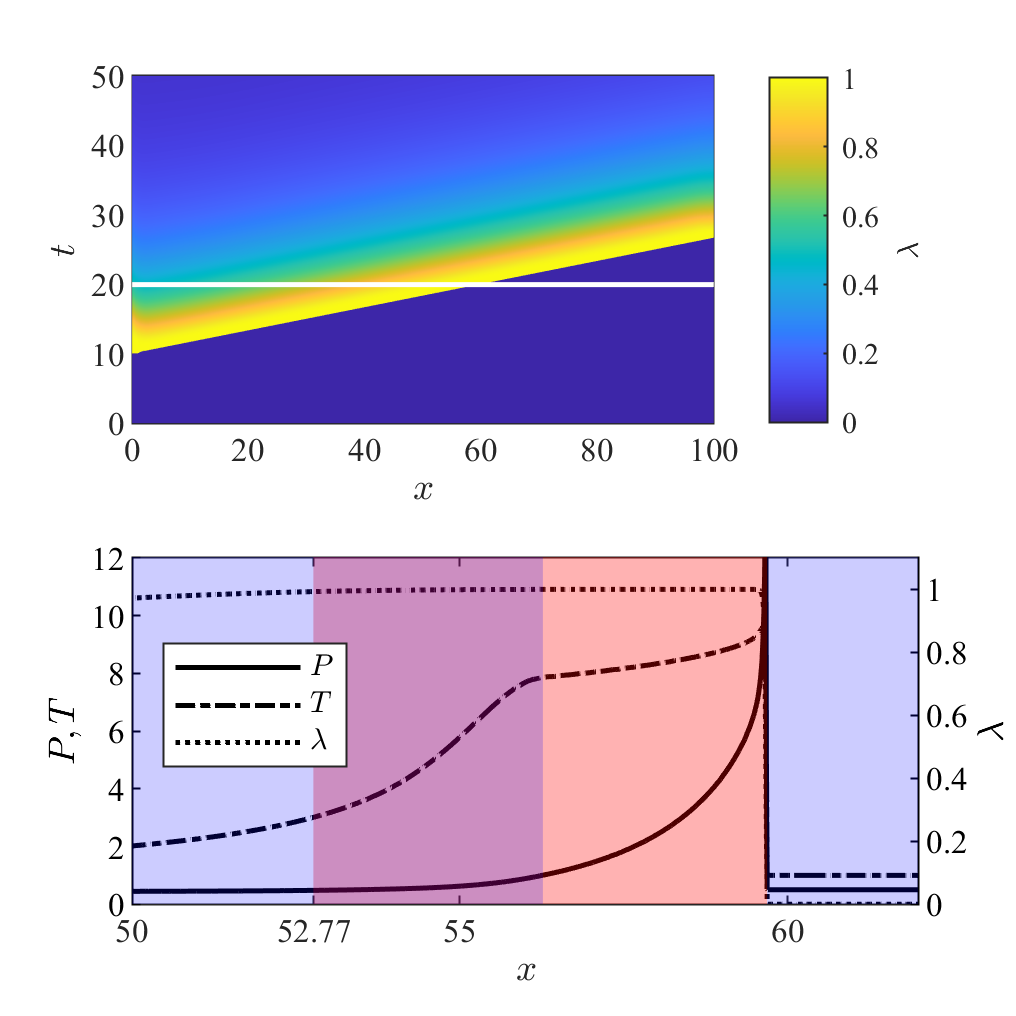}
        \put(13,47){(b)}
        \put(13,94.5){(a)}   
        \end{overpic}
        \caption{Detonation tube profile for slow propellant regeneration: $A_r = 0.5$, $E_a = 12.5$, $\beta = 0.1$, and $Da=10$.}
		\label{fig:solitary}
\end{figure}

The numerical detonation tube experiments of Section \ref{sec:validation} are repeated, though with all model source terms active. Because there now exists mass and energy inflow and outflow, a stabilization period of $0<t<10$ lets the domain reach a steady state before initiating chemical reactions. A combustion impulse is applied at $t=10$ from $0<x<3$ for model systems with slow ($\beta=0.1$) and fast ($\beta=0.4$) propellant regeneration, all else constant.  Figure \ref{fig:solitary} details the waveform and space-time history of the resultant detonation wave for the case with slow propellant regeneration. Consistent with simulations from the lumped-volume impulse response and the detonation tube validation survey, a solitary detonative pulse forms and travels through the domain. On the tail end of the wave, there is no accumulation of energy that could lead to steady deflagration or the formation of multiple waves. The waveform consists of several distinct regions. In Fig. \ref{fig:solitary}a, the detonation travels into the quiescent medium at the operating level prescribed by the input/output balance of the lumped-volume model. The shock-reaction structure leads a region, shaded in red, where $H(P)=0$. No regeneration of propellant or inlet mass flux occurs in this region. The region shaded in purple denotes the region where propellant regeneration co-exists with combustor temperatures that exceed $T = T_{ign.}$. Note that $\omega>0$ always, but for illustrative purposes we show this region as an indication of the potential for the accumulation of energy within the domain. For $\beta=0.1$, this overlap does not lead to an accumulation of energy within the domain and the state decays to its initial, ambient condition. 

\begin{figure}[t]
        \centering
        \begin{overpic}[width=1\linewidth]{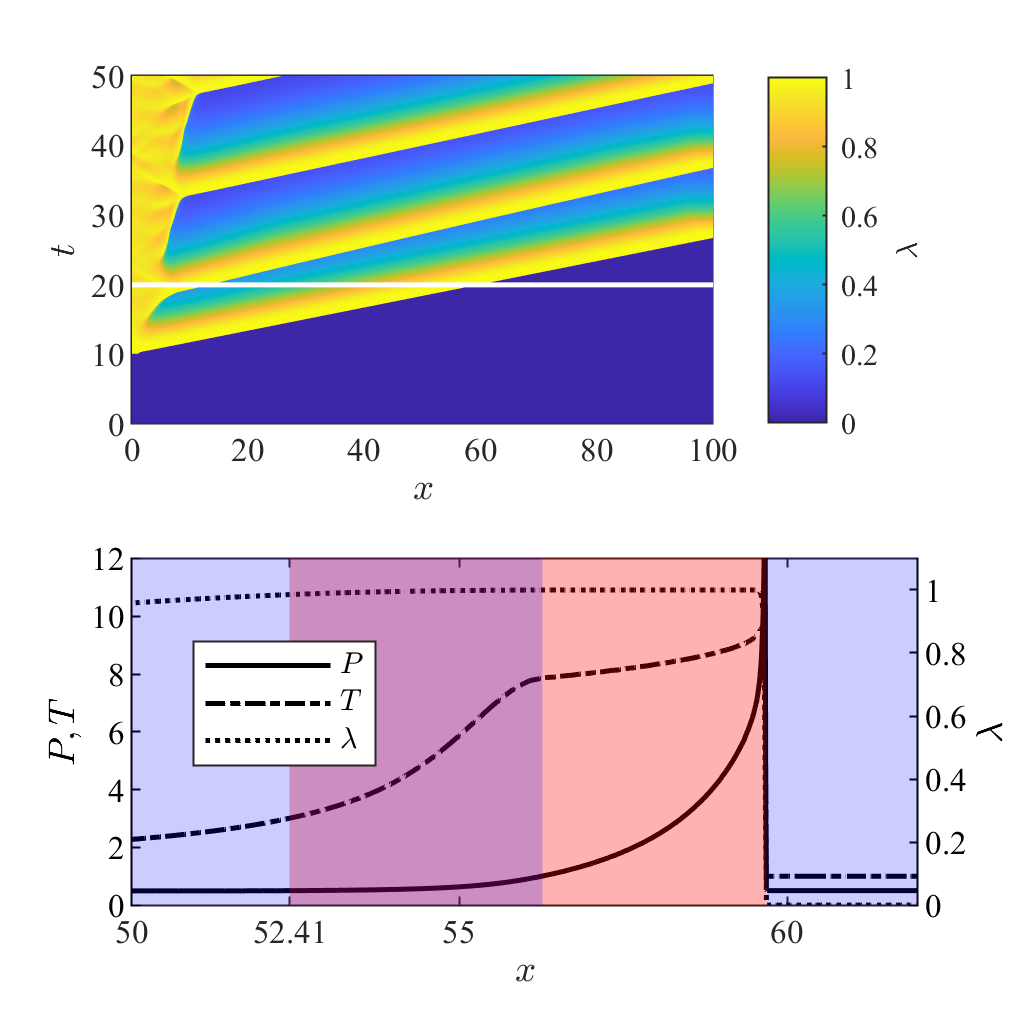}
        \put(13,47){(b)}
        \put(13,94.5){(a)}   
	    \end{overpic}
        \caption{Detonation tube profile for fast propellant regeneration: $A_r = 0.5$, $E_a = 12.5$, $\beta = 0.2$, and $Da=10$.}
		\label{fig:wavetrain}
\end{figure}

Increasing $\beta$ leads to increasing the amount of time and distance in which newly introduced propellant co-exists with high combustor temperatures. This is exhibited in Fig. \ref{fig:wavetrain}, where the conditions are such that there is accumulation of energy from parasitic deflagration on the tail-end of the detonation wave. Note (i) the increased length of the purple shaded region over that of $\beta=0.1$ and (ii) the temperature at $x=50$ for $\beta=0.1$ versus $\beta=0.2$: that of the latter case is about 10\% higher. This energy accumulation self-accelerates, eventually forming a wavetrain behind the leading detonation wave.  Each peak behind this leading wave can either transition to a separate detonation - in which case it becomes acoustically isolated from the preceding wave - or remain entrained behind the prior wave and exist as a deflagration bump traveling at an identical speed. In this specific case, a wavetrain of detonations traveling at identical speeds follow the leading detonation wave. Complex wall-contact surface interactions lead to wave counterpropagation near $x=0$.

\subsection{Periodic Domain}
In this section the model system (Eq. \ref{eq:pde}) is applied to a 1-D periodic line of length $L=30$ with activation energy $E_a=10$. Note that in choosing this relatively low activation energy, expected is an increase of parasitic deflagration. We do this to exacerbate the effects of parasitic deflagration on wave dynamics and thermodynamic trends. A comprehensive review of modes of operation and behavior with respect to model parameters is outside of the scope of this paper. Instead, presented are a subset of numerical experiments whose properties are representative of the model. With the length of the domain and the chemical kinetic model constants fixed, we perform a set of simulations sweeping through values of $\beta$ and $Da$, effectively changing the propellant mixing and convective time scales, and $A_r$. The overall area ratio changes the operating pressure level of the combustor and therefore the degree of combustor-injector interactions. Similarly, the blockage parameter $c$ changes the recharge and expulsion time scales, though for this study we hold $c=0$. A summary of model parameters used for this section is given in Table \ref{tab:params1D}.

\begin{figure*}[t]
        \centering
        \begin{overpic}[width=1\linewidth]{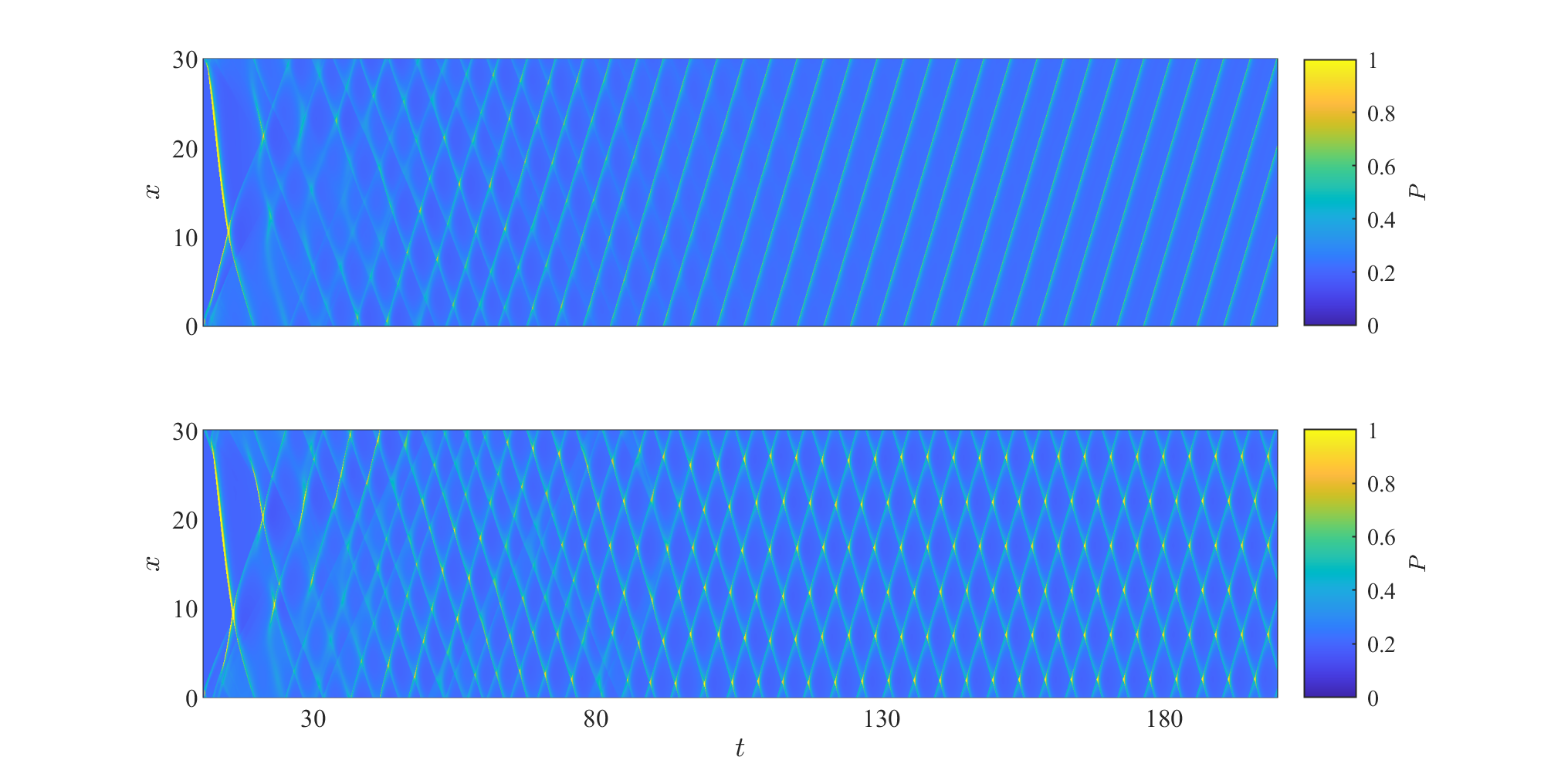}
        \put(13,47){(a)}
        \put(13,23.5){(b)}    
        \end{overpic}
        \caption{(a) Start-up transient and mode-locking of three detonation waves for $A_r = 0.2$, $\beta = 0.075$, and $Da = 20.4$. Compare with the experimental spatiotemporal dynamics of Fig. \ref{fig:expStartup}. (b) Start-up transient and mode-locking for counter-propagating waves for $A_r = 0.2$, $\beta = 0.116$, and $Da = 10.0$. Compare with the experimental spatiotemporal dynamics of Fig. \ref{fig:intro}.}
		\label{fig:startup}
\end{figure*}

\begin{table}
	\caption{1-D Periodic Model Parameters}
	\label{tab:params1D}
	\centering
	\begin{tabular}{ccccccccccc}
	\hline
	$\gamma$ &$q$& $L$ & $E_a$ & $T_{ign.}$ & $c$\\
	
	1.3 & 25 & 30 & 10 & 3 & 0\\

	\hline
	\end{tabular}
\end{table}

\begin{figure}[t]
        \centering
        \begin{overpic}[width=1\linewidth]{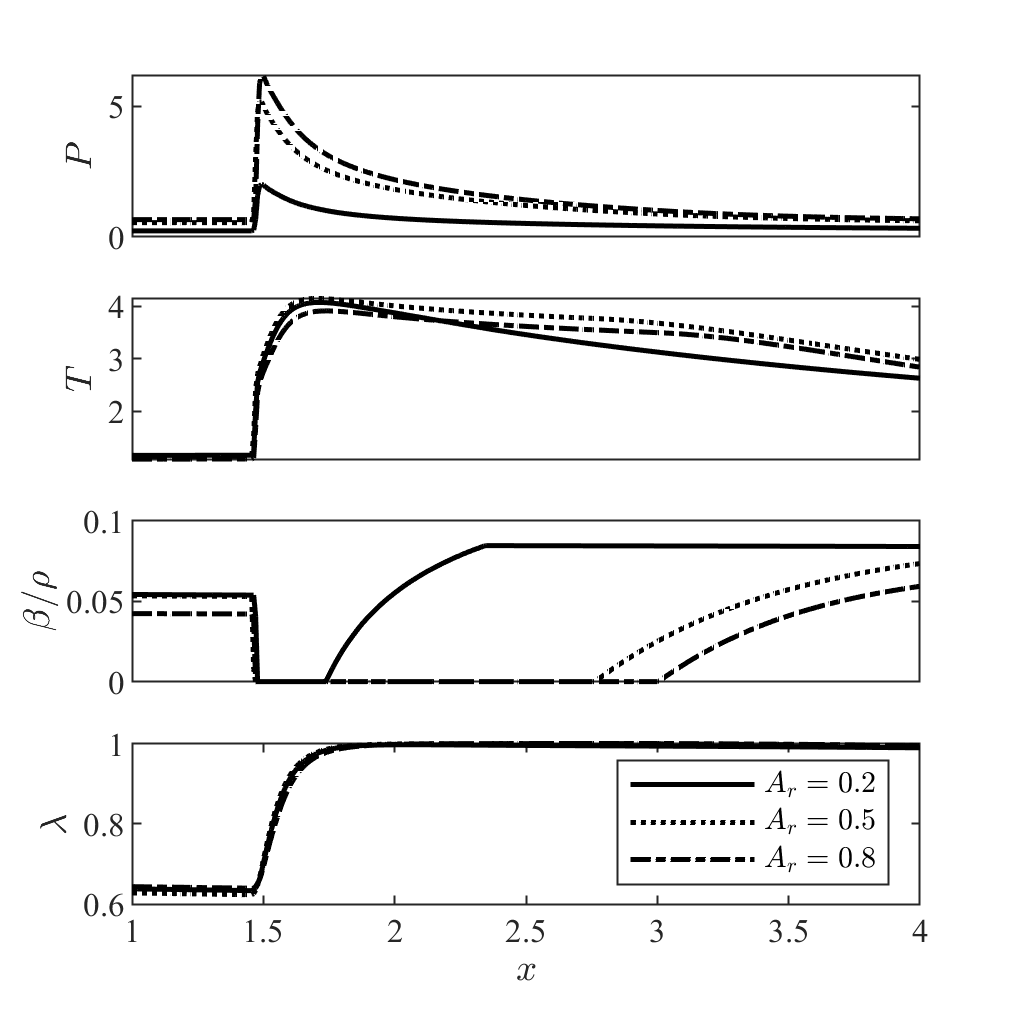}
        \put(13,95){(a)}
        \put(13,72.5){(b)}    
        \put(13,50.5){(c)}    
        \put(13,29){(d)}
	    \end{overpic}
        \caption{A close-up view of the detonation wave profiles for various area ratios. Profiles for large area ratio simulations exhibit a greater refractory period and less parasitic deflagration. Compare with the decaying profiles of Fig. \ref{fig:AR_variation}.}
		\label{fig:compare}
\end{figure}

A representative model start-up transient is shown in Fig. \ref{fig:startup}a. A sinusoidal profile of $\lambda(x,t<10) = \frac{1}{2}sin\left( \frac{2\pi x}{L}\right) + \frac{1}{2}$ is used as the initial condition from which reactions are started. At $t=10$, $\beta$ is switched from zero to the desired value for the simulation. The region $0<{x}<3$ is simultaneously hit with an artificially high pre-exponential factor for several time steps to begin chemical reactions. The asymmetry in $\lambda$ promotes quick transition to a final steady-state. This initialization procedure is applied to all numerical experiments in this section. In the case of Fig. \ref{fig:startup}a, with $A_r=0.2$, $\beta = 0.075$, and $Da=20.4$, the steady state is three co-rotating waves after a period of wave counter-propagation. However, the propagation direction is opposite that of the initiated wave. Compare the spatiotemporal dynamics of Fig. \ref{fig:startup}a with those of Fig. \ref{fig:expStartup}.

To compare the detonation front structure across different area ratios, three simulations were performed at a common value of $\beta$ and $Da$ that resulted in a single propagating wave. These simulations were performed at $A_r$ of 0.2, 0.5, and 0.8 for $\beta=0.085$ and $Da=10$ with no blockage ($c=0$). Resultant profiles are shown in Fig. \ref{fig:compare}. As in the simulations of the lumped-volume model (Fig. \ref{fig:AR_variation}), higher area ratio simulations lead to a greater refractory period behind the wave front during which only exhaustion and rarefaction, not injection, can occur. For low area ratios, such as the $A_r=0.2$ case of Fig. \ref{fig:compare}, propellant can be introduced very soon after the passage of the detonation wave, despite the high temperature present during injection. Thus, the lower area ratio cases can be more susceptible to parasitic deflagration than the high area ratio counterparts.

\begin{figure}[t]
        \centering
        \begin{overpic}[width=1\linewidth]{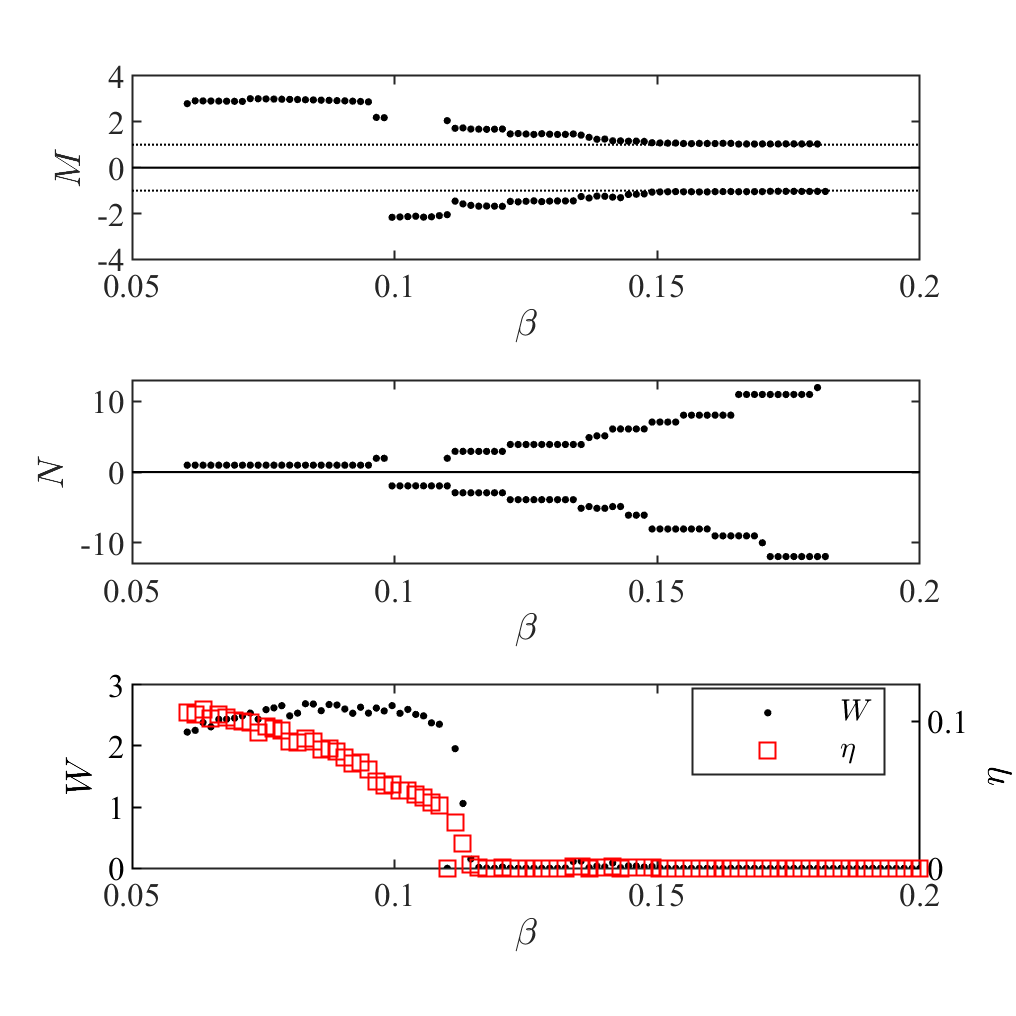}
        \put(13,95){(a)}
        \put(13,65){(b)}    
        \put(13,35){(c)}    
	    \end{overpic}
        \caption{A sweep through values of $\beta$ that result in traveling waves for $A_r=0.2$ and $Da=10$. In (a), the traveling wave Mach number is displayed, where a negative Mach corresponds to a clockwise direction of travel. Similarly, in (b), the number of waves, $N$, is given with positive values denoting counter-clockwise rotation and negative values denoting clockwise rotation. As $\beta$ increases, wave co-rotation ceases in favor of stable counter-propagation. In the limit of large $\beta$, the wave Mach numbers decay to the acoustic velocity of the combustion chamber and then cease to exist altogether, merging into a planar deflagration front.}
		\label{fig:bif}
\end{figure}

A set of simulations sweeping through $\beta$ was performed for a low area ratio ($A_r=0.2$) and holding the Damk\"ohler number constant at $Da=10$.  A summary of the sweep is shown in Fig. \ref{fig:bif}, displayed as traveling wave Mach number, $M = v_{wave}/\sqrt{\gamma T}$, through the bifurcation parameter $\beta$ for the range of values that result in traveling waves. The region of greatest wave Mach is for the single traveling wave case. The Mach number reaches a maximum at $M = 3.0$ at $\beta = 0.0725$. Note that the theoretical maximum steady Mach number, the Chapman-Jouguet Mach number for the self-sustained detonation, $M_{{CJ}^+}$, evaluates to $M_{{CJ}^+} = 5.339$. Thus, for the cases presented in Fig. \ref{fig:bif}, the highest Mach number wave is 56\% of the theoretical value for the propellant. As the value of $\beta$ is increased, the system crosses a point of criticality where colliding waves can continue to propagate after collision. This is caused by both the short refractory period associated with lower area ratios and also the quickening of the time scale for propellant mixing. In Fig. \ref{fig:bif}, this transition occurs at $\beta=0.11$ with the system moving from two co-rotating waves to four waves; two moving clockwise and two counter-clockwise in a regular fashion. Further increases in $\beta$ lead to a dramatic increase in the number of waves. At $\beta=0.18$, 12 counter-clockwise and 12 clockwise waves were present, though they travel at the acoustic velocity of the medium. Beyond $\beta=0.18$, the waves merge into a planar deflagration front. The bifurcation diagrams for $A_r$ of 0.5 and 0.8 are qualitatively similar. 

\begin{figure}[t]
        \centering
        \begin{overpic}[width=1\linewidth]{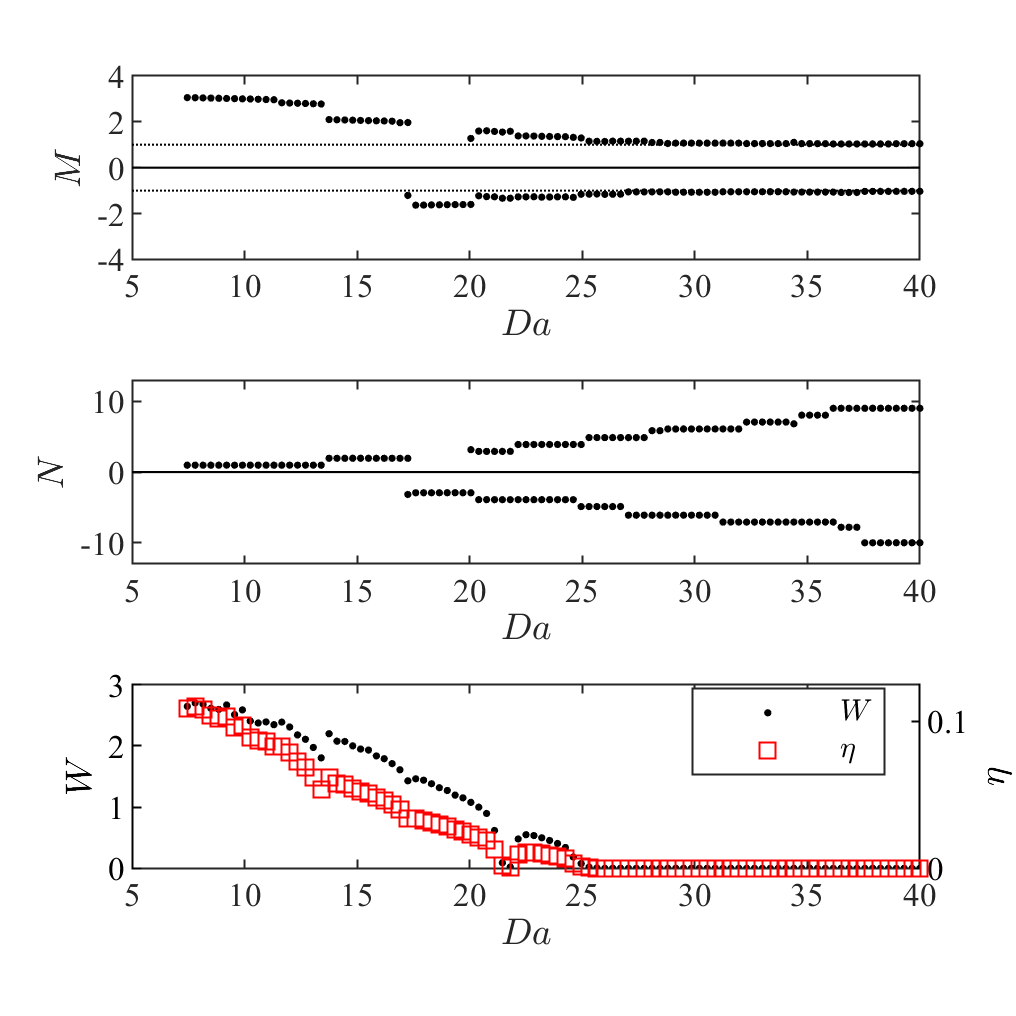}
        \put(13,95){(a)}
        \put(13,65){(b)}    
        \put(13,35){(c)}    
	    \end{overpic}
        \caption{A sweep through values of $Da$ that result in traveling waves for $A_r=0.2$ and $\beta=0.075$. As was seen in the case of large $\beta$, in the limit of large $Da$, the wave Mach numbers decay to the acoustic velocity of the combustion chamber and then cease to exist, merging into a planar deflagration front.}
		\label{fig:bifDa}
\end{figure}

An analogous sweep of $Da$ is performed holding all other parameters constant ($A_r = 0.2$, $\beta = 0.075$, and $c=0$). The bifurcation diagrams are shown in Fig. \ref{fig:bifDa}. A similar progression of co-to-counter- wave propagation occurs, though in keeping $\beta$ low, more co-rotating waves are found to stably exist (up to three for this domain length). After developing 9 clockwise and 10 counter-clockwise waves with $Da=40$, the waves merge into a planar deflagration.

In increasing $\beta$, $Da$, and/or $c$ the result is an increase in parasitic deflagration, though through different physical mechanisms. In increasing $\beta$, propellant is reintroduced into the domain quickly after the passage of a detonation wave, even if the domain temperature is high enough to activate the chemical kinetics. This is analogous to the waveform of Fig. \ref{fig:wavetrain}. In the case of a high Damk\"ohler number, the chemical time scale is faster than the convective time scale of the fluid. Thus, integrated heat release through combustion for any period of time is higher than a low Damk\"ohler counterpart. On the tail of the decaying detonation waves, the refractory period dictated by the area ratio and $\beta$ remain steady, but once injection is activated, the balance of injection and combustion begins to favor combustion with increasing $Da$. Lastly, by constricting the exit of the combustor, the amount of time required to purge the exhaust gas and to refill the chamber increases. Thus, newly introduced propellant is in contact with hot exhaust gases for longer periods of time with increasing $c$, leading to parasitic deflagration. In the extreme, if $c=1$, there is no outflow - the system behaves as a bomb calorimeter.

\begin{figure}[t]
        \centering
        \begin{overpic}[width=1\linewidth]{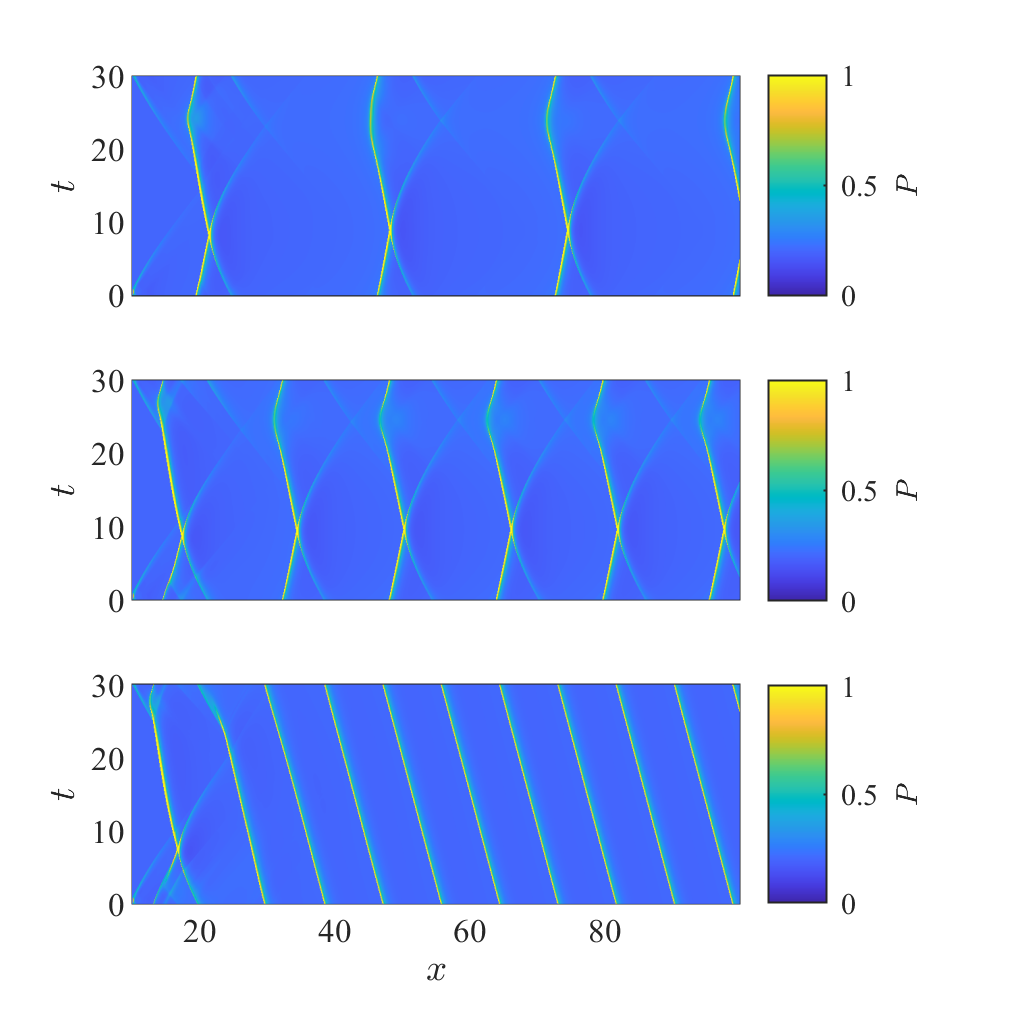}
        \put(13,95){(a)}
        \put(13,65){(b)}    
        \put(13,35){(c)}    
	    \end{overpic}
        \caption{Prior to being able to support continuously traveling waves, periodic nucleation-extinction wave pair structures persist through time. Simulations correspond to $\beta=0.075$, $A_r=0.2$, and $Da$ of 5, 6.75, and 7.8 (top to bottom).}
		\label{fig:slapping}
\end{figure}

Before proceeding, we additionally present the qualitative transition from simulations that cannot support traveling waves to those that can. Figure \ref{fig:slapping} shows this transition as a series of simulations with incremental changes in the model Damk\"ohler number. With a low $Da$, a continuous traveling wave is not supported. However, combustion and injection can still oscillate in the form of plane waves with period longer than that of a traveling wave. In the limit of low $Da$, these plane waves behave as the state oscillations shown in Fig. \ref{fig:EA_variation}. As $Da$ increases, asymmetry in the plane waves (as induced by the asymmetric initial condition) can be seen, eventually resembling the head-on collision of a pair of counter-propagating waves. Because $Da$ is still too low to support continuous propagation, the waves consume all available propellant and are quickly dissipate. `Hot spots' still exist in the chamber and serve as nucleation points for a new wave pair and the process repeats. This is analogous to `slapping' modes presented by many in literature \citep{Anand2019,Anderson2020}. The same qualitative transition exists in ramping up $\beta$ from an initially low value.

\subsection{Thermodynamics}

\begin{figure}[t]
        \centering
        \begin{overpic}[width=1\linewidth]{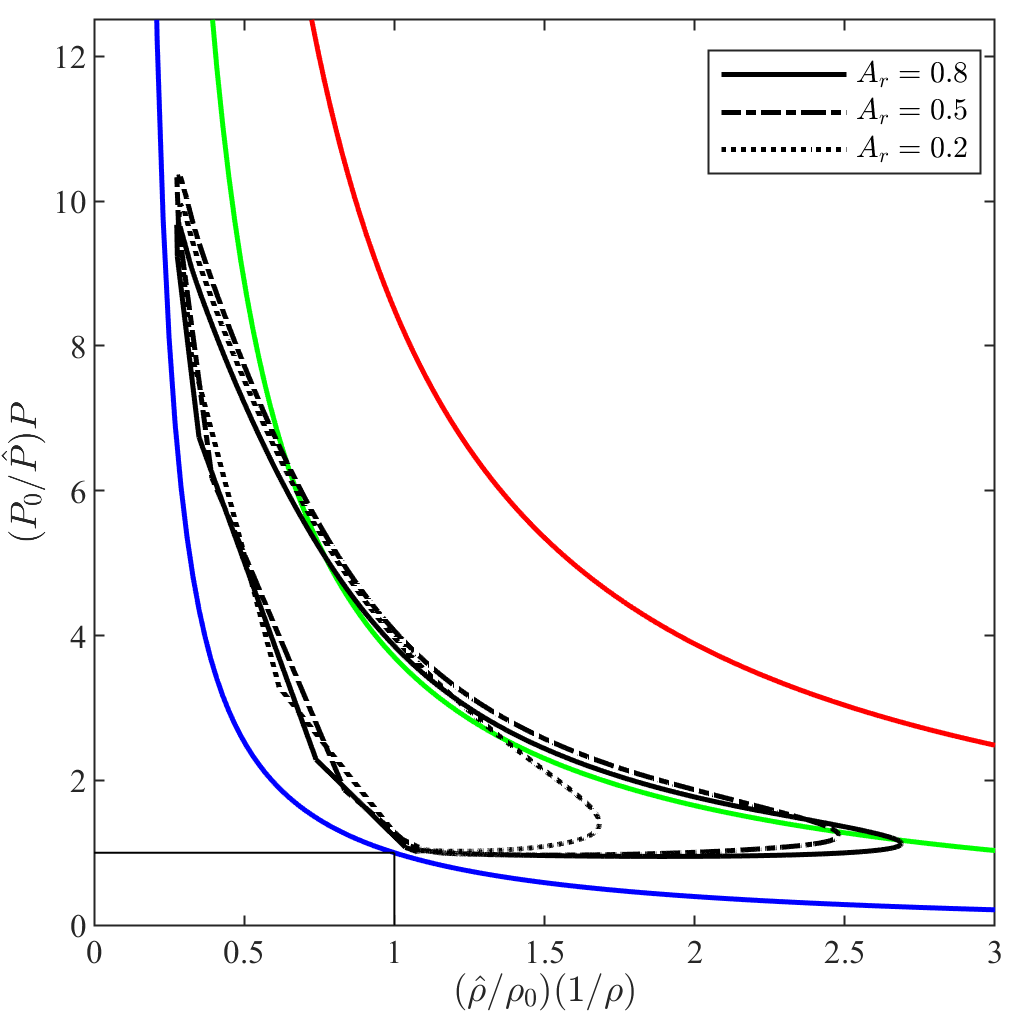}
        \put(17,20){$q=0$}
        \put(26,65){$q=15.9$}    
        \put(45,61){$q=25$}    
	    \end{overpic}
        \caption{Thermodynamic power cycles corresponding to the numerical experiments of Fig. \ref{fig:compare}. $\beta = 0.085$ and $Da=10$ for all cycles. Three Hugoniot curves are overlaid corresponding to different heat release values: $q=0$, $q=15.9$, and $q=25$. The heat release associated with perfectly mixed propellant is $q=25$.}
		\label{fig:cycles}
\end{figure}

We refer to the detonation profiles shown in Fig. \ref{fig:compare}. Because the traveling detonation waves are self-sustained in the model, the power cycle can be plotted for each of these cases in $P-1/\rho$ and $T-dS$ coordinates. These power cycles are given in Fig. \ref{fig:cycles} alongside several Hugoniot curves. The enclosed area of the $P$-$1/\rho$ curves correspond to the work output per cycle of the system. For $A_r=0.2$, this non-dimensional work output is 2.87, while those of $A_r=0.5$ and $A_r = 0.2$ are 3.84 and 3.65, respectively. Dividing the values of work output by the round-trip time of the wave gives a representative power output for the cycle. The non-dimensional output power is 0.17, 0.46, and 0.42 for $A_r=0.2$, 0.5, and 0.8, respectively. The three Hugoniots correspond to the adiabatic shock curve ($q=0$), the heat associated with the maximum change in $\lambda$ for the $A_r=0.8$ case ($q=15.9$), and the heat associated with completely mixed propellant ($q=25$). The three presented cycles are qualitatively similar. Each begins near the thermodynamic origin of $(P,1/\rho) = (1,1)$ (corrected by the base state of the combustor for each $A_r$, $\hat{P}$ and $\hat{\rho}$, as given by Fig. \ref{fig:fixedPts}). After shock compression and heating, combustion adds heat to the fluid along a Rayleigh line until the propellant is depleted. For these cases, this approximately occurs at the $q=15.9$ Hugoniot. Expansion occurs in the vicinity of this Hugoniot back to near-rest values of pressure. A near-constant pressure injection process resets the state to the thermodynamic origin.

\begin{figure}[t]
        \centering
        \begin{overpic}[width=1\linewidth]{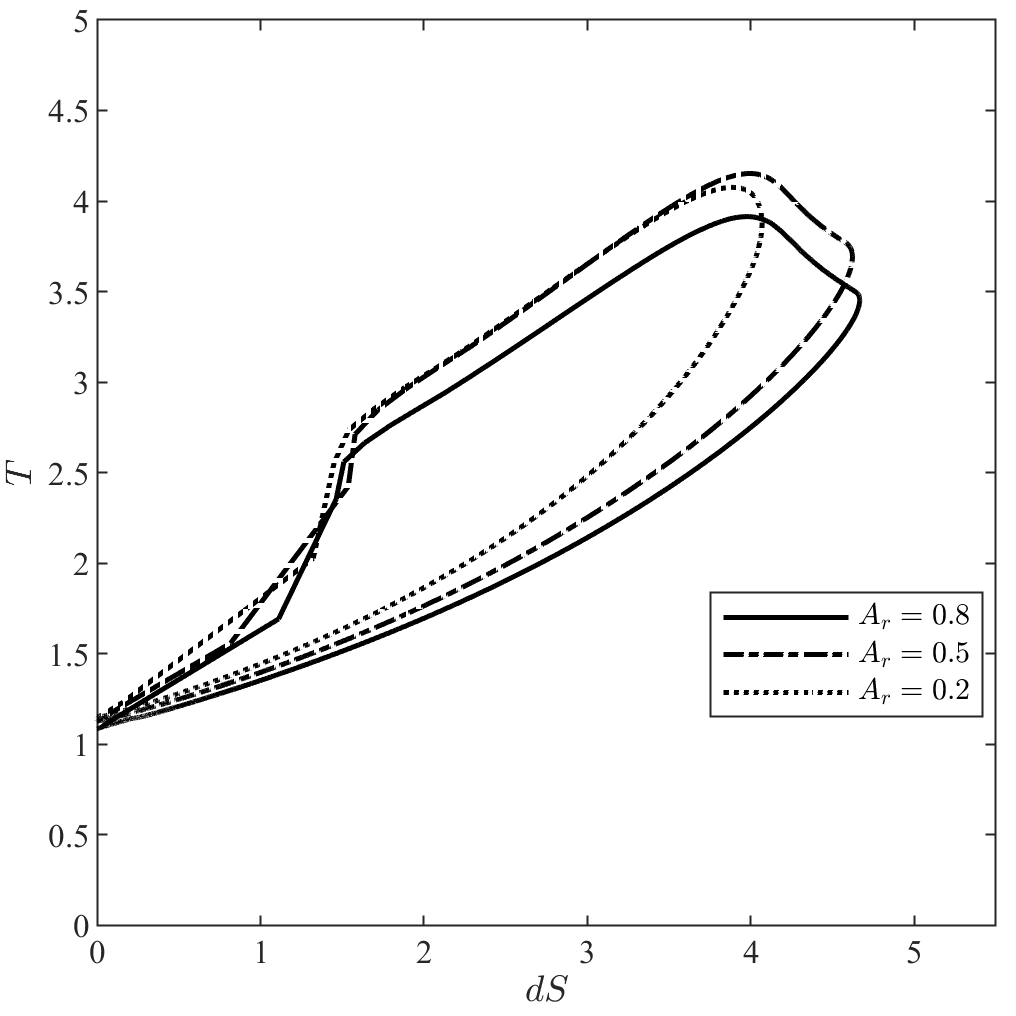}
	    \end{overpic}
        \caption{Temperature-entropy diagrams for thermodynamic cycles corresponding to the numerical experiments of Fig. \ref{fig:compare} and the power cycles of Fig. \ref{fig:cycles}. $\beta = 0.085$ and $Da=10$ for all cycles.}
		\label{fig:TS}
\end{figure}

Thermodynamic cycle efficiency is given by $\eta_{cycle}=W/Q_{in}$ where $Q_{in}$ is the enclosed area of the cycle in the $T-dS$ diagram. This heat input for $A_r=0.2$, 0.5, and 0.8 is 3.18, 4.20, and 3.98 with cycle efficiencies of 0.90, 0.91, and 0.92. However, this metric is very misleading: the heat input as indicated by the closed trajectory in the $T-ds$ diagram is not equivalent to the heat input to the system over one cycle. To illustrate this point, consider a planar deflagration throughout the domain - no area is enclosed in either the $P-1/\rho$ or $T-dS$ diagrams. The amount and rate of heat addition exactly offset those of dissipation (in the form of exhausting to an ambient condition). Offsetting combustion-exhaustion is not captured in the $T-dS$ diagrams of Fig. \ref{fig:TS}. Therefore, we introduce an appropriate representative efficiency metric:

\begin{equation}
    \eta_{overall} = \frac{W}{q\int_{0}^{L} \frac{\omega}{\rho} dx } .
\end{equation}

The metric $\eta_{overall}$ relates the available mechanical work output with the total integrated heat release over the domain, thereby including all parasitic deflagration that may occur in the domain. Note that this integration occurs over the 1-D domain versus tracking an individual fluid particle through time. For steady co-rotating waves of the presented model, all fluid particles undergo the same thermodynamic trajectory. Thus, because the cycle is closed in both space and time, the integrals are equivalent. For the three cases in \ref{fig:cycles}, this efficiency evaluates to 9\%, 11\%, and 13\% for $A_r=0.2$, 0.5, and 0.8, respectively. These values are the manifestation of the imposed time scales and chemistry model for these specific cases - they should not be interpreted as representative performance metrics for RDEs in general. Figures \ref{fig:bif}c and \ref{fig:bifDa}c detail the available work output and thermal efficiency through sweeps of $\beta$ and $Da$.

\begin{figure}[t]
        \centering
        \begin{overpic}[width=1\linewidth]{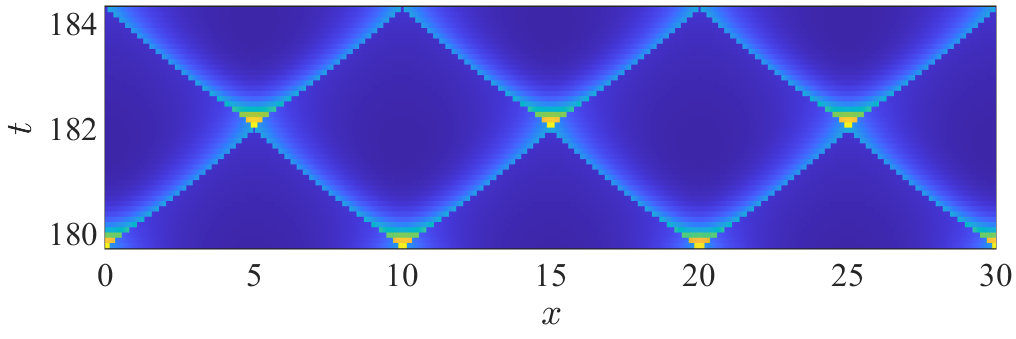}
	    \end{overpic}
        \caption{The minimum repeating time period for wave counter-propagation is shown. A representative work and power output is obtained by temporally integrating the work performed on a distribution of fluid particles throughout the domain.}
		\label{fig:counterPropPower}
\end{figure}

For cases exhibiting wave counter-propagation, like that of Fig. \ref{fig:startup}b, there does not exist a single representative power cycle as there is for cases with co-rotating waves. This is because the spatial distribution of the state (pressure, density, and energy) for counter-propagation does not have a continuous symmetry associated with the traveling waves; i.e., a representative wave profile cannot be ``frozen'' in space by imposing an offsetting advection speed as in the case with co-rotating waves. However, a work and power output can still be numerically computed. In Fig. \ref{fig:counterPropPower}, the minimum repeating period of wave counter-propagation  for the case of $A_r=0.2$, $\beta=0.116$, and $Da=10$ is shown. To compare this case with those of Fig. \ref{fig:cycles}, we temporally integrate the work done to a representative distribution of fluid particles over this period. For this condition, the net work output for each minimum repeating time period over the entire domain is 0.56 and the power output is 0.12. As the base-to-peak amplitude of the detonation waves diminish (for example, as $\beta$ or $Da$ are increased, corresponding to moving right in Figures \ref{fig:bif} and \ref{fig:bifDa}), the available mechanical work output similarly decreases. In the limit of weak wave propagation, the combustion wave fronts become acoustic in nature with no appreciable pressure rise. The thermodynamic cycle shrinks from a well-defined path that encircles a large area of $P$-$1/\rho$ space to a single, isolated point (a planar deflagration) from which no mechanical work can be extracted.

\section{Discussion} \label{sec:discussion}

In this article, we relate the injection, mixing, exhaustion, and combustion processes present in the rotating detonation engine in a simplified modeling framework. Injection, mixing, and exhaustion are modeled with a zero-dimensional lumped-volume approach that is linked to the Euler equations of an invscid, compressible fluid on a periodic line. The resulting model system provides the necessary energy and mass in- and out-flow pathways needed to nucleate and shape coherent combustion wave fronts. For certain parameter regimes, the model produces self-organized states of traveling and standing waves. These states include persistent co-rotation of an integer number of waves (clockwise or counter-clockwise). Furthermore, through wave counter-propagation, the formation of regular patterns has been recovered.

In this section, highlighted are the key takeaways of this study. First is the implicit relationship of time and spatial scales associated with injection, mixing, exhaustion, and combustion. Second is the set of mechanisms that lead to wave counter-propagation.

\subsection{Time and Spatial Scales}
The nondimensionalization of Section \ref{sec:nondim} yielded three dimensionless groupings: $\alpha$, $\beta$, and $Da$. $\alpha$ relates the mass flux of the reference (injection) condition through the injection area versus the annular area. $\beta$ relates the convective time scale of the fluid ($t_{0} = z_0/u_0$) to the imposed time scale for the mixing process. Lastly, the Damk\"ohler number relates the convective time scale of the fluid with the chemical time scale of the reactant. In the presented simulations, these time scales are separated by multiple orders of magnitude, with the time scale of combustion being the shortest and the mixing time scale being the longest. However, note that these two time scales in particular are intimately coupled through an additional time scale: that of a combustion wave traversing the periodic domain. Naturally, the transit time of the wave is directly related to its speed, which is governed by classical detonation theory. Required is a sufficient refill and mixing of propellant in the traveling detonation wave's path to sustain the velocity of the wave. In this manner, the time scale for mixing (and injection, though in general it is that of mixing that is more restrictive) must be comparable to the period of the wave. The waves undergo a `self-adjustment' process to accommodate the mismatch of these scales. Supposing the wave period to be shorter than what is required for propellant mixing, the waves will decay in strength (amplitude) and speed. Decreasing wave speed allows for longer mixing periods and, in general, lower values of $\lambda$. This now presents a potential surplus of propellant through which the waves can accelerate, now reducing the mixing time. Depending on model parameters, this `self-adjustment' process can lead to mode-locking, steady wave modulation, chaotic propagation, or transitions to a different number of waves. 

\begin{figure}[t]
        \centering
        \begin{overpic}[width=1\linewidth]{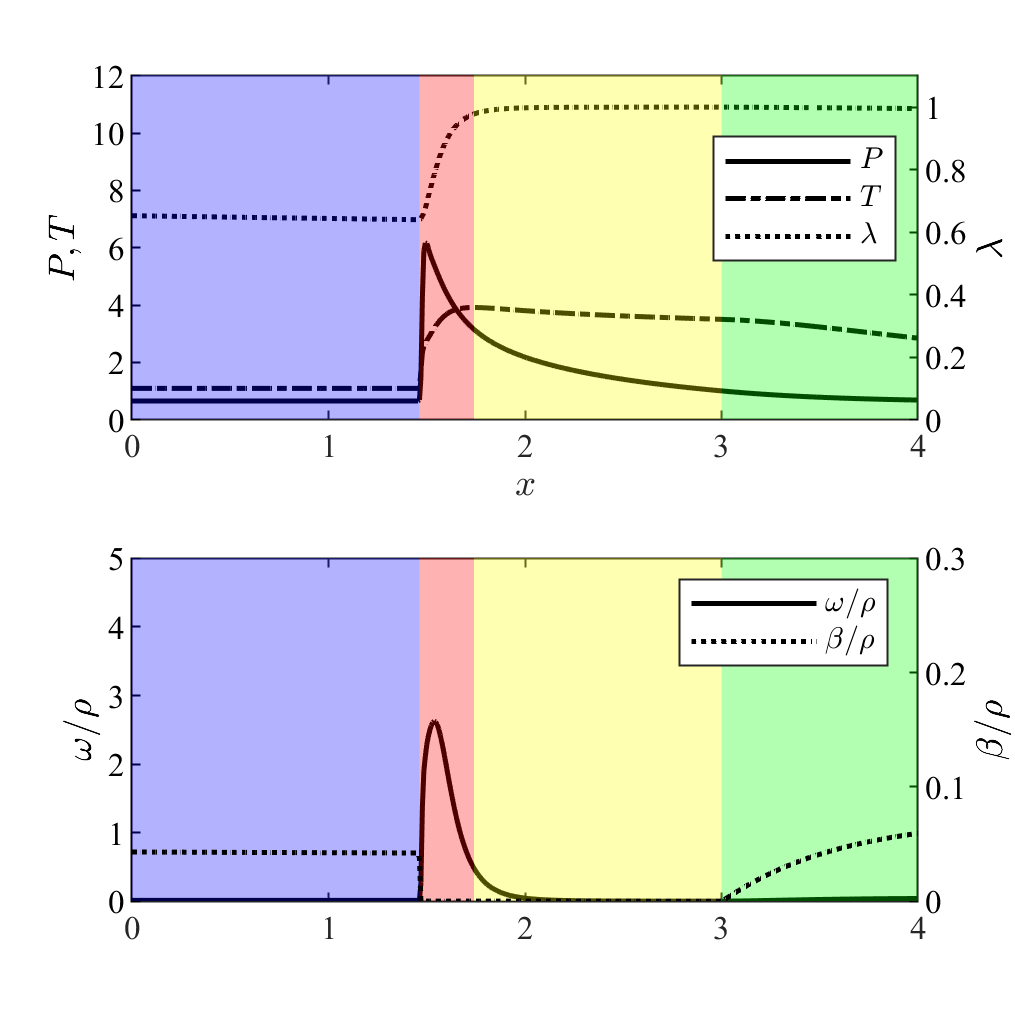}
        \put(13,95){(a)}
        \put(13,47){(b)}   
        \put(22,20){Mixing}
        \put(16,36){Combustion}
        \put(37,36.5){\vector(1,0){6.5}}
        \put(53,20){Exhaust}
        \put(73.5,20){Injection}
        
	    \end{overpic}
        \caption{The different shaded regions correspond to the activation or deactivation of source terms in the model system. The result is a waveform governed by different combinations of active source terms that give distinct localities of different balance physics. In this simulation, the high pressure wave temporarily blocks the injection and mixing processes, providing a refractory period before injection can occur. In red, the physics are combustion-dominant. In yellow, with $\lambda\approx 1$, exhaust processes are dominant. Injection and mixing occur through the rest of the domain. Note that this is a zoomed-in view of the detonation wave front and does not include the full domain. $A_r=0.8$, $\beta=0.085$, and $Da=10$.}
		\label{fig:spatialScales}
\end{figure}
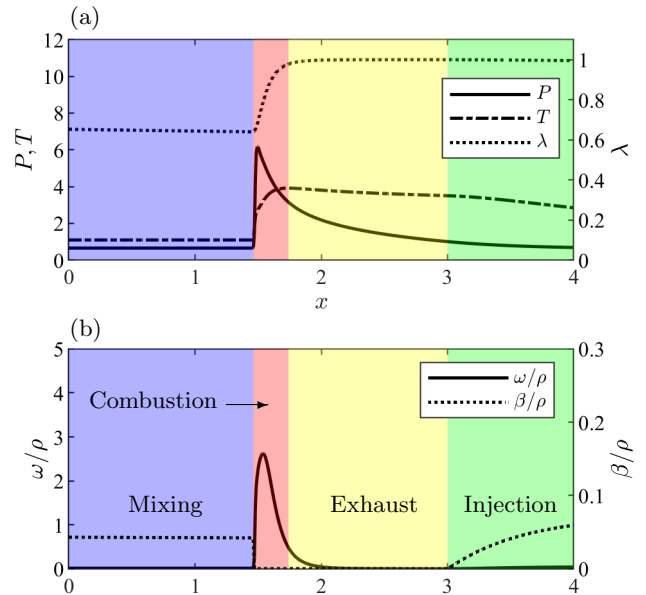

The separation of time scales in the RDE give a similar set of spatial scales where the local dominant balance physics can vary drastically from those of the overall system. To exemplify this we examine an example waveform from a single wave simulation shown in Fig. \ref{fig:spatialScales} and remind the reader of the source term (Eq. \ref{eq:source}) presented in Section \ref{sec:model} and reproduced here for convenience:

\begin{equation}
\mathbf{{S}} = 
\begin{bmatrix}
\alpha\left( A^+ H({P}) - A^- \sqrt{{P}{\rho}}\right)
\\
0
\\
\frac{\alpha}{\gamma - 1}\left(A^+ H({P}) - {T} A^- \sqrt{{P}{\rho}}\right) + {\omega} {q}
\\
\widetilde{\omega} + {\rho}{\beta} H({P}) \lambda + \alpha\left( A^+ H({P}) - A^-\sqrt{{P}{\rho}}\right)\lambda
\end{bmatrix} ,
\end{equation}

In the zoomed-in view of the wave front in Fig. \ref{fig:spatialScales}, one can see the separation of physical processes (and their respective scales) that shape the time histories of the state variables. Beginning with the shock front, combustion is initiated and occurs quickly relative to the other processes. The shock/reaction structure occurs in the red shaded region over the span of about 0.25 spacial units. In this region, the local physics is dominated by chemical reactions (terms containing $\omega$). After depletion of reactant, the local physics switch to exhaustion-dominated, indicated by yellow shading. With $\lambda$ near or at unity, the reaction source terms evaluate to near-zero values. Likewise, $H(P)$ is zero until $P$ decays to values below one. Therefore, the only terms contributing to the physics in $\mathbf{S}$ are the exhaustion terms (those containing $\sqrt{{P}{\rho}}$). This region occurs over the span of approximately 1.25 spacial units. This constitutes the refractory space and period behind the traveling waves. Once $P$ falls below 1, $H(P)$ is activated on a linear ramp until $P$ is low enough to induce injector choking (when $H(P) = 1$). In Fig. \ref{fig:spatialScales}, this is indicated in green. In this region, $\lambda$ is still near unity, discouraging chemical reactions (note that $\beta$ is low enough to similarly discourage parasitic deflagration, unlike that of Fig. \ref{fig:wavetrain}). The source terms active in this region are those containing both $H(P)$ and $\sqrt{{P}{\rho}}$. This distinct region ends at the point where the mass flow into and out of the domain balance - on the order of 10 spatial units for the presented case. Once balanced, the terms containing  $H(P)$ and $\sqrt{{P}{\rho}}$ exactly offset, leaving the propellant mixing term ($\rho \beta H(P) \lambda)$ as the sole driver of dynamics in the model.  This region is indicated by blue shading and makes up the remainder of the domain. Note that $\lambda$ is no longer near unity and is actively approaching zero. Chemical reactions are free to proceed as $(1-\lambda) > 0$, but in this region for this case, the temperature is too low to activate the kinetics. Thus, only propellant mixing contributes to the dynamics in this region. Note that these distinct regions correspond to the major features of the thermodynamic cycles presented in Fig. \ref{fig:cycles}. The shock-reaction structure is the compression and heat input portion of the cycle. Cycle closure is provided by exhaustion and refill as prescribed by the source terms in $\mathbf{S}$.

For cases where significant propellant mixing co-exists with high chamber temperatures, exacerbated parasitic deflagration may occur and lead to a \textit{merger of all of the shaded regions}. First, the injection (green) and mixing (blue) regions overlap and can lead to an elevated base temperature in the domain and therefore a non-zero contribution from chemical reactions. In this scenario, terms containing $\omega$ and the propellant mixing term ($\rho \beta H(P) \lambda)$ are simultaneously active. In the extreme, the chemical reactions accelerate to the point at which the traveling waves can no longer be sustained (see Fig. \ref{fig:bif} for high values of $\beta$). The result is a planar deflagration front where \textit{all} of the terms in $\mathbf{S}$ are active. As the resultant structure is a plane wave, all spatial derivatives are zero and the system reduces to the coupled set of ODEs given in Eq. \ref{eq:ode} with properties shown in Figs. \ref{fig:AR_variation} and \ref{fig:EA_variation}. 

\subsection{Counter-propagation}

\begin{figure}[t]
        \centering
        \begin{overpic}[width=1\linewidth]{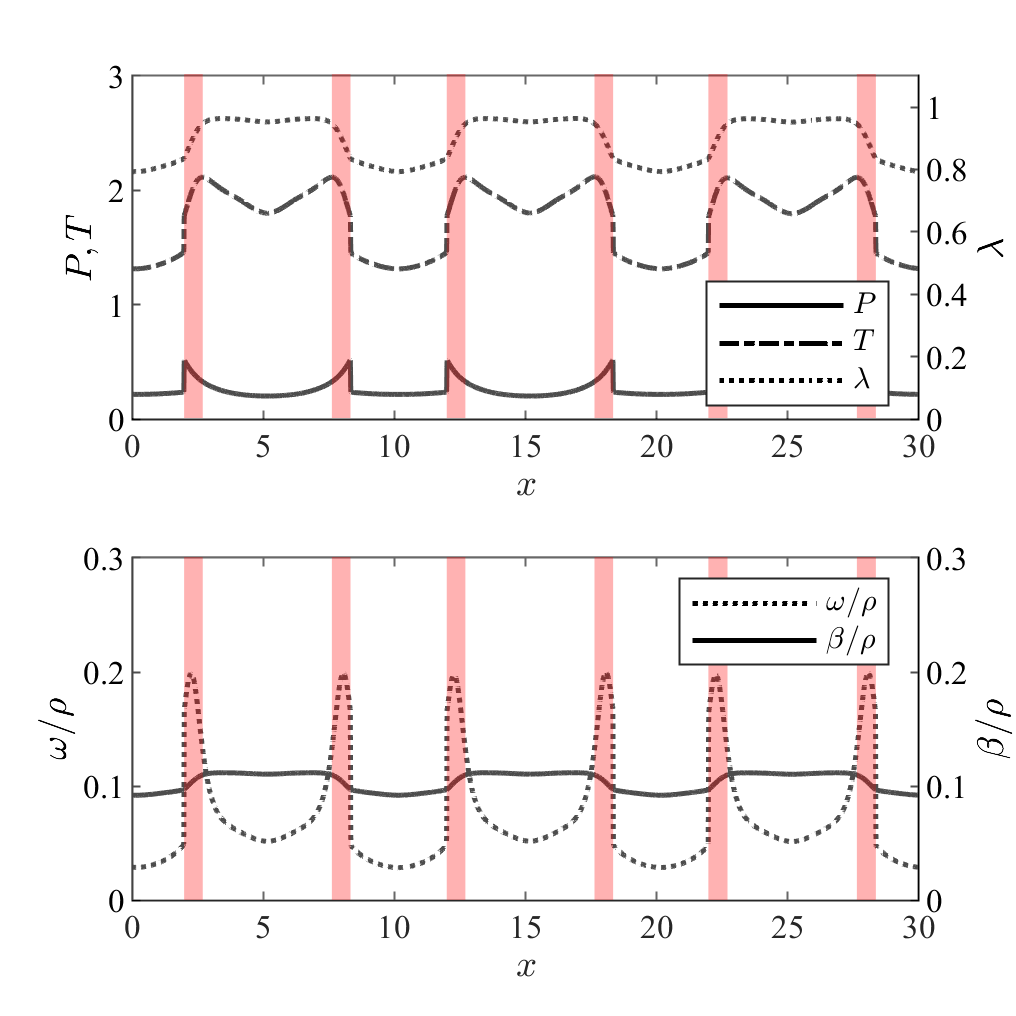}
        \put(13,95){(a)}
        \put(13,47){(b)}  
	    \end{overpic}
        \caption{The lack of an effective refractory period allows for stable wave counter-propagation such as in this simulation with $A_r=0.2$, $\beta=0.116$, and $Da=10$. There are two distinct dominant regions: combustion-dominant in red and not combustion-dominant (everywhere else). All source terms are active throughout the domain.}
		\label{fig:spatialScalesCounterProp}
\end{figure}

The ability of two waves to continue to propagate after collision is related to the strengths of the waves and the length of the refractory period following the shock fronts. In cases with short refractory periods, caused by a lack of deactivation of $H(P)$, colliding waves can continue to propagate if sufficient injection and mixing occur on the tail ends of the waves. For long refractory periods this becomes impossible: the waves after collision will dissipate as no injection or mixing has occurred (see: Fig. \ref{fig:slapping}). A subtle connection to the injection-to-annulus area ratio is made: for low $A_r$ models, $H(P)$ may never be deactivated as the peak pressure associated with the traveling waves is lower than that of injection. Because $H(P)$ remains activated, little to no injector-feedback induced refractory period exists. The only effective refractory period is provided by where $\lambda$ is near unity, limiting the progress of combustion. The resulting scenario is similar to that of high $\beta$ values: propellant mixing and high temperatures may co-exist, leading to the both the elimination of the injection-induced refractory period and the merger of injection and mixing spatial regions. Figure \ref{fig:spatialScalesCounterProp} shows the waveform for a wave counter-propagation case with $A_r = 0.2$, $\beta = 0.116$, and $Da = 10.0$. $H(P)$ is active everywhere - only two distinct regions exist in the model: combustion-dominant and not combustion-dominant. In the region not dominated by combustion, the physical effects of all source terms are approximately of the same order, including combustion. All processes progress through time, though attainment of a steady state (planar deflagration, in this case) is unattainable. The high wave count quickly sends the local state far-from-equilibrium. The combination of the lack of refractory period and elevated $\beta$ allows waves to collide and reorganize indefinitely. Note that in this case, the reaction thickness is longer than those of the high area ratio and co-rotating wave cases. Furthermore, the reactions are \textit{incomplete} for this case - $\lambda$ reaches a maximum at $0.94$. This represents the value (and spatial location) at which the rate of combustion is equal to the rate of reintroduction of propellant. The \textit{effective} refractory period provided by complete combustion ($\lambda=1$) no longer exists but at the points of wave collision: injection, mixing, exhaustion, and combustion occur everywhere.

\section{Conclusion} \label{sec:conclusion}
Motivated by experimental observations of regular spatiotemporal patterns formed by rotating detonation waves, we formulate a tractable model that blends approaches from both RDE analog models and reactive Euler equations of inviscid compressible fluid flow with the goal of replicating the observed spatiotemporal dynamics. The presented model is a hybrid lumped-volume/CFD system that treats injection, mixing, and exhaustion as processes acting on a lumped volume. A version of the 1-D reactive Euler equations for an inviscid compressible flow are used to simulate the gasdynamics along the annulus of an RDE. The two are linked through source terms in the reactive Euler equations. 

Three dimensionless groups are found to govern the system behavior: $\alpha$ - the ratio of mass flow through the injection area versus the annular area for a reference condition, $\beta$ - the ratio of the convective time scale of the fluid to the time scale for propellant mixing, and the Damkohler number, which relates the convective time scale of the fluid to the chemical reaction time scale. Successful detonation wave propagation can occur if the time scales of the physical processes related by these dimensionless groups are separated by several orders of magnitude. 

When properly separated, the local balance physics of the RDE rotate through periods (in time) or regions (in space) of shock-induced combustion, rapid exhaustion, propellant injection, and finally propellant mixing. These processes trace out a thermodynamic power cycle similar from which available work output and power can be extracted. Supposing the balance provided by the separation of spatial and time scales is altered, stable wave counter-propagation can be sustained. Thus, the pattern formation as observed in experiments can be recovered by this model. However, we find that wave-counter-propagation corresponds to a reduction in work output per cycle and an increase of parasitic deflagration. In the extreme, the counter-propagating waves increase in number until, for the parameter regime investigated in this article, dozens co-exist in the chamber traveling at the acoustic velocity of the medium. Further increase of energy flux causes the catastrophic transition to a steady planar deflagration front.

\section*{Acknowledgements}
This work was supported in part by the US Air Force Center of Excellence on Multi-Fidelity Modeling of Rocket Combustor Dynamics award FA9550-17-1-0195. JNK acknowledges support from the Air Force Office of Scientific Research (AFOSR) grant FA9550-17-1-0329. Experiments were performed at the University of Washington High Enthalpy Flow Laboratory.

\bibliographystyle{ieeetr}
\bibliography{prf}

\end{document}